\documentclass[aps,groupedaddress,showpacs,letterpaper,twocolumn,reprint,nofootinbib,longbibliography]{revtex4-1}
\usepackage[colorlinks=true,citecolor=blue]{hyperref}
\usepackage{graphicx}
\usepackage{braket}
\usepackage{physics}
\usepackage{amsmath}
\usepackage{amssymb}
\usepackage{amsthm}
\usepackage{stackrel}
\usepackage{comment}
\usepackage{todonotes}
\usepackage{color}
\usepackage{dsfont}
\usepackage{mathrsfs}
\usepackage{bm}
\usepackage{units}
\usepackage{algorithm} 
\usepackage[noend]{algpseudocode}
\usepackage{tikz}
\usetikzlibrary{quantikz}

\newtheorem{definition}{Definition}

\renewcommand{\braket}[1]{\langle #1 \rangle}
\usepackage{upgreek}
\usepackage[capitalise]{cleveref}
\crefformat{section}{\S#2#1#3}
\crefname{condition}{condition}{conditions}

\usepackage{bbold}
\usepackage[normalem]{ulem} 
\usepackage{academicons}
\definecolor{orcidlogocol}{HTML}{A6CE39}

\newcommand\rotpi{\rotatebox[origin=c]{180}{$\pi$}}

\begin{document}

\title{Two-qubit gate in neutral atoms using transitionless quantum driving}
\author{Archismita Dalal}
\email{archismita.dalal1@ucalgary.ca}
\affiliation{Institute for Quantum Science and Technology, University of Calgary, Alberta T2N~1N4, Canada}
\author{Barry C.~Sanders}
\email{sandersb@ucalgary.ca}
\affiliation{Institute for Quantum Science and Technology,
	University of Calgary, Alberta T2N~1N4, Canada}
\date{\today}

\begin{abstract}
A neutral-atom system serves as a promising platform for realizing gate-based quantum computing because of its capability to trap and control several atomic qubits in different geometries and the ability to perform strong, long-range interactions between qubits;
however, the two-qubit entangling gate fidelity lags behind competing platforms such as superconducting systems and trapped ions. 
The aim of our work is to design a fast, robust, high-fidelity controlled-Z (CZ) gate, based on the Rydberg-blockade mechanism, for neutral atoms. 
We propose a gate procedure that relies on simultaneous and transitionless quantum driving of a pair of atoms using broadband lasers.
By simulating a system of two interacting C\ae sium atoms, including spontaneous emission from excited levels and parameter fluctuations, we yield a Rydberg-blockade CZ gate with fidelity 0.9985 over an operation time of $0.12~\upmu$s. 
Our gate procedure delivers CZ gates that are superior than the state-of-the-art experimental CZ gate and the simulated CZ gates based on adiabatic driving of atoms.
Our results show that our gate procedure carries significant potential for achieving scalable quantum computing using neutral atoms.
\end{abstract}
\maketitle

\section{Introduction}
The aim of this work is to develop a procedure for implementing a high-performance entangling gate on a pair of neutral alkali atoms trapped in an optical lattice.
Our work is important because neutral-atom quantum computing is one of the most promising architecture of quantum computing~\cite{Saffman16,APS19,HBS+20,MW21,Shi22}.
Popular two-qubit gates for atomic qubits utilize the strong dipole-dipole interaction between Rydberg-excited atoms~\cite{BBL16}, and such ``Rydberg-blockade" gates have been explored both in theory~\cite{JCZ+00,LFC+01,Saffman16,Shi22} and experiments~\cite{LKS+19,MCS+20,GSS+22}.
Gate performance is typically quantified in terms of the Bell-state preparation fidelity~($F_\text g$), which is indicative of the quantum-algorithmic performance, and gate-implementation time~($T_\text g$), which is required to be short for avoiding decoherence and yielding high clock rates~\cite{NC02}.
We propose a procedure for designing time-dependent functions for laser pulses driving the pair of atoms simultaneously and thereby effecting the controlled-Z~(CZ) gate.
In the presence of spontaneous emission from excited levels of these atoms and major technical imperfections, our procedure results in a simulated Rydberg-blockade CZ gate with higher $F_\text g$ and lower $T_\text g$ than state-of-the-art CZ implementations~\cite{LKS+19,GSS+22}.

The promise of neutral atoms for quantum computing materialized with the advent of Rydberg-blockade gate procedures~\cite{JCZ+00} and the capability to trap multiple atoms almost deterministically using optical tweezers~\cite{BDL+16,EBK+16}.
The potential of this platform stems from its inherent ability to control several qubits coherently in different geometries~\cite{XLM+15, WKWW16} 
and to facilitate long-range interactions between qubits.
Quantum information is encoded either in the long-lived hyperfine levels of the ground state (``ground-ground''~\cite{MW21}) or in a superposition of ground and highly-excited Rydberg states (``ground-Rydberg''~\cite{MW21}), where all qubits are identical and can be well isolated from their environment.
Experiments have already achieved $F_\text g>0.99$ (`two nines')
for single-qubit gates~\cite{XLM+15,WKWW16}, accompanied by high efficiencies for state initialization and detection~\cite{MCS+20}.
In addition to performing universal computation, this platform has the ability to simulate quantum dynamics~\cite{SLK+21,EWL+21}.
Industrial efforts towards achieving commercial quantum computing is rapidly intensifying~\cite{coldquanta,quera,pasqal,atomcomputing}.

Although neutral atoms deliver a versatile platform for quantum technologies, scalable quantum computing using ground-ground encoding suffers from low $F_\text g$
(below two nines)
for two-qubit gates~\cite{LKS+19, GSS+22}.
In contrast, 
ground-Rydberg encoding delivers two-nine fidelity but suffers from low coherence~\cite{MCS+20}.
Neutral-atom two-qubit gate fidelities are far behind competing platforms such as superconducting systems~\cite{BKM+14,KWS+21} and ion-traps~\cite{WBD+19}, and they do not even reach the required thresholds for error-correcting codes~\cite{Saffman16}.
One promising method for delivering `three-nine' two-qubit gate fidelity~($F_\text g>0.999$) is to employ adiabatic pulses with optimal parameters~\cite{SBD+20}, but such gates are fundamentally slow.
Here, we propose a best-of-both-worlds approach using shortcut-to-adiabaticity~\cite{GRK+19} for designing pulses that yield fast, robust and high-fidelity CZ gates for atomic qubits.

We adapt the transitionless quantum driving (TQD) technique~\cite{Berry09}, commonly used for achieving shortcut-to-adiabaticity, to design time-dependent laser pulse sequences that yield a high-performing Rydberg-blockade gate.
We consider both dipole, i.e.\ one-photon, and quadrupole, i.e.\ two-photon, driving of atoms between their ground and Rydberg levels.
Dipole driving is challenging because it requires an ultraviolet laser, which is extremely sensitive to Doppler broadening~\cite{HJP+14}.
Quadrupole driving circumvents these problems but requires stronger driving and yields lower $F_\text g$ for adiabatic CZ gates~\cite{SBD+20}.
Our novel method for pulse design applies both dipole and quadrupole cases
and combines our modified TQD technique with a pulse-concatenation technique~\cite{LKS+19}
to yield fast, high-fidelity two-qubit gates. 

We numerically integrate the quantum master equation to simulate the Rydberg-blockade CZ gate on C\ae sium~(Cs) atoms including spontaneous emission and major technical imperfections.
Our analysis predicts $F_\text g\approx0.998$ over $T_\text g=0.12~\upmu$s for dipole-driving, whereas,
for quadrupole driving, we predict $F_\text g\approx0.975$ over $T_\text g=0.24~\upmu$s.
Our proposed gates outperform the recent experimental CZ gate on Cs atoms, for which $F_\text g\approx0.955$ over $T_\text g=0.8~\upmu$s~\cite{GSS+22}.  
Our procedure delivers gates that are at least twice as fast as the adiabatic CZ gates~\cite{SBD+20}, while keeping $F_\text g$ and maximum laser intensities almost equal.
Moreover, to achieve a noiseless (closed-system) fidelity of 0.99, laser intensity required by the adiabatic CZ gate increases five times faster than our CZ gate for decreasing $T_\text{g}$.
Thus, our gate procedure is potentially superior to existing alternatives and paves the way for scalable neutral-atom quantum computing.

Our paper is organised as follows. 
Our background in~\S\ref{sec:background} describes CZ-gate implementation on atomic qubits, followed by proposals for adiabatic CZ gates and techniques to characterise these gates in simulations.
Additionally, in~\S\ref{sec:background}.
we elaborate on pertinent background for TQD.
We then explain our approach,
in~\S\ref{sec:approach},
for computing time-dependent Rabi-frequency and detuning functions for laser pulses, which,
when applied to a pair of Cs atoms,
yield fast, high-fidelity CZ gates.
In~\S\ref{sec:results}, we derive expressions for these pulse functions and present numerical results for performances of our gates compared with those obtained by the adiabatic procedure~\cite{SBD+20}.
Then we analyse our results and their implications in~\S\ref{sec:discussion} and conclude in~\S\ref{sec:conclusion}.

\section{Background}
\label{sec:background}
In this section, we explain key concepts and methods necessary for designing and characterizing a two-qubit gate in neutral atoms.
We begin by elaborating the pertinent background on implementing, both in experiment~(\S\ref{subsec:experiment}) and in simulation~(\S\ref{subsec:adiabaticgate}), CZ gates on atomic qubits using the Rydberg-blockade phenomena.
Then we discuss numerical techniques for simulating such gates in the presence of spontaneous emission and parameter fluctuations~(\S\ref{subsec:characterize}).
Finally, we explain transitionless quantum driving and its application to quantum control~(\S\ref{subsec:TQD}).
\begin{table}[]
    \centering
    \begin{tabular}{|c|c|}
    \hline
    $F_\text g$ & Bell-state preparation fidelity or gate fidelity\\
    \hline
    $T_\text g$ & gate time\\
    \hline
    CZ& controlled-Z gate \\
    \hline
    CZ$_\phi$& \cref{eq:CZphimatrix} (CZ$_0$ = CZ) \\
    \hline
    & subscript denoting label for a laser pulse\\
    $\ell$ & ($\ell=0,\text{B,R}$ for pulses driving transitions $\ket1\leftrightarrow\ket{\text r}$, \\
     & $\ket1\leftrightarrow\ket{\text p}$ and $\ket{\text p}\leftrightarrow\ket{\text r}$, respectively.)\\
    \hline
     & superscript denoting label for an excitation model\\
    e & (e=d for the one-photon or dipole case, \\
    & and e=q for the two-photon or quadrupole case)\\
    \hline
    $\sigma_{x,y,z}$ & localization parameters of the atoms\\
    \hline
    $\sigma^\text{i}_\ell$ & fluctuation in laser intensity \\
    \hline
    $\sigma^\text{D,e}$ & dephasing rate due to Doppler effect\\
    \hline
    $\sigma^\text{m,e}$ & dephasing rate due to magnetic field fluctuation\\
    \hline
    $\sigma^-_F$ & standard error of estimating fidelity~$F$ \\
    \hline
    \end{tabular}
    \caption{List of frequently used symbols.}
    \label{tab:symbols}
\end{table}

\subsection{Implementation of CZ gates}
\label{subsec:experiment}
Here we describe state-of-the-art CZ implementations on ground-ground qubits of alkali atoms. 
Our focus are CZ gates that are effected by the Rydberg-blockade interaction and are thus native to Rydberg atoms,
which means this gate is implemented in the system in a natural way, and is universal. 
We begin by elaborating on the Rydberg-blockade CZ gate and the ``standard procedure" for realizing this gate. 
Then we explain the gate procedure used for achieving state-of-the-art CZ gates on alkali atoms.
Finally, we summarize the experimental implementation on Cs atoms and state the typical magnitudes of experimental parameters, which will be used in our numerical simulation.

Among different techniques for constructing two-qubit gates~\cite{SWM10,Shi22}, Rydberg-blockade gates are the most popular because their fidelities are insensitive to the exact magnitude of the dipole coupling strength and are less susceptible to external motional degrees of freedom~\cite{JCZ+00}.
The dipole-dipole interaction between two Rydberg atoms, which is quantified by the blockade shift~$B$~\cite{WS08}, comes into effect only for an initial two-qubit state~$\ket{11}$ and shifts its phase when this interaction is stronger than the driving lasers.
This ``Rydberg-blockade condition" leads to a negligible excitation of the two-atom state~$\ket{\text{rr}}$ throughout the gate implementation.
The strong Rydberg-Rydberg interaction yields a CZ gate, up to some single-qubit phase~\cite{LKS+19,GKG+19}, which is robust against the variation in $B$ and consequently in the interatomic separation.

The standard procedure for implementing a fast Rydberg-blockade gate employs a sequential driving of the two atoms~\cite{JCZ+00}.
The procedure commences with preparing two atoms in a superposition of the two-qubit computational basis states $\{\ket{00}, \ket{01}, \ket{10}, \ket{11}\}$, where one atom is treated as control and the other as target. 
The transition between $\ket1$ and $\ket{\text{r}}$ of each atom is effected by individual addressing with on-resonant narrow-band optical pulses.
The procedure involves the following sequence: $\pi$ pulse on control atom, followed by a $2\pi$ pulse on target atom and finally another $\pi$ pulse on control atom. This sequence, in the ideal case, yields
\begin{equation}
\label{eq:CZmatrix}
    \operatorname{CZ} := \operatorname{diag}(1,1,1,-1),
\end{equation}
up to a global phase $\pi$.
The most recent implementation of the CZ gate on Cs atoms using this procedure yields $F_\text g=0.89$ over $T_\text g=1.12~\upmu$s~\cite{GKG+19}.

As opposed to this standard procedure, the state-of-the-art procedure for implementing the CZ gate is based on simultaneous excitation of both atoms~\cite{LKS+19}.
This procedure is developed by identifying sectors, generated by block-diagonalizing the Hamiltonian matrix, in the unitary dynamics of the system corresponding to the four computational basis states.  
The $\ket1\leftrightarrow \ket{\text r}$ transition of both atoms is driven through a two-photon process of global off-resonant narrow-band optical pulses.
By choosing appropriate pulse parameters, a $2\pi$-pulse effects a full Rabi cycle for $\ket{11}$, but partial cycles for $\ket{01}$ and $\ket{10}$.
A sequence of two such identical $2\pi$-pulses, with an appropriate relative phase shift of~$\phi_R$, yields a controlled-phase gate
\begin{equation}
\label{eq:CZphimatrix}
    \operatorname{CZ}_\phi:= \operatorname{diag}\left(1,\mathrm{e}^{\mathrm{i}\phi},\mathrm{e}^{\mathrm{i}\phi},\mathrm{e}^{\mathrm{i}(2\phi+\pi)}\right),
\end{equation}
which is equal to the standard CZ gate~\eqref{eq:CZmatrix} up to a single-qubit phase $\phi$. 
Based on this procedure, recent experiments yield $F_\text g\ge0.974$ for Rubidium atoms~\cite{LKS+19} and $F_\text g\sim0.955$ for Cs atoms~\cite{GSS+22}.

In the state-of-the-art experiment~\cite{GSS+22},
which we elaborate here in detail,
cold $^{133}$Cs atoms at temperature $T_\text{a}=5~\upmu$K are trapped in a $7\times7$ site optical lattice with period $d=3~\upmu$m. 
At this temperature, the localization parameters of the atoms in the transverse (on the lattice) and axial (perpendicular to the lattice) directions of the trap are $\sigma_x=\sigma_y=0.14~\upmu$m and $\sigma_z=0.75~\upmu$m, respectively.
A qubit is encoded in the hyperfine clock states of the atom, i.e.
\begin{align}
\label{eq:qubit}
\ket0 :=& \ket{6 \text{s}_{\nicefrac12},f=3,m_f=0},
    \nonumber\\
\ket1 :=& \ket{6 \text{s}_{\nicefrac12},f=4,m_f=0},
\end{align}
with a frequency separation of 9.2~GHz, and $T_1=4$~s and $T_2^*=3.5$~ms.
The transition between $\ket1$ and the Rydberg state~$\ket{\text{r}}=\ket{75 \text{s}_{\nicefrac12},m_j=-\nicefrac12}$ is carried out using a two-photon excitation via an intermediate state
\begin{equation}
\label{eq:intermedstate}
\ket{\text{p}}:=\ket{7 \text{p}_{\nicefrac12}},
\end{equation}
with a radiative lifetime of $\tau_{7 \text{p}_{\nicefrac12}}=0.155~\upmu$s.

A CZ$_\phi$ gate~\eqref{eq:CZphimatrix} based on the simultaneous-driving procedure is implemented on a pair of atoms separated by $3d$.  
Each atom is irradiated by two lasers, which are focussed on the atom using acousto-optic deflectors.
The off-resonant two-photon transition between $\ket1$ and $\ket{\text r}$ is realized using a blue-detuned (B) $\sigma_+$ laser of wavelength $\lambda_\text B=459$~nm for $\ket{1}\leftrightarrow\ket{\text{p}}$ and a red-detuned (R) $\sigma_-$ laser of wavelength $\lambda_\text R=1040$~nm for $\ket{\text{p}}\leftrightarrow\ket{\text{r}}$. 
We denote laser pulses for two- and one-photon excitation by $\ell=\text{B,R}$, and $\ell=0$, respectively; see~\cref{tab:symbols}.
The detuning from $\ket{\text p}$ is high ($\approx760$~MHz) to reduce scattering.
These two beams counter-propagate along the trap's axial direction with an effective wave-vector
\begin{equation}
k^\text{q}_\text{eff}
:=\nicefrac{2\pi}{\lambda_\text B}-\nicefrac{2\pi}{\lambda_\text R}
=7.64~\upmu \text{m}^{-1}.
\end{equation}
where the superscript `q' denotes quadrupole driving (\cref{tab:symbols}).
The beam waist ($\nicefrac1{\text{e}^2}$ intensity radius) of each laser is $w_{\text B(\text R)}=3~\upmu$m, which results in negligible Rabi frequencies at the neighbouring sites.
The effective Rabi frequency for each atom is 1.7~MHz and $\nicefrac{B}{2\pi}=3$~MHz for the atom-pair, which violates the Rydberg-blockade condition.

The performance of the gate is measured in terms of $F_\text g$ and $T_\text g$.
All atoms in the trap are initially prepared in state~$\ket1$ by optical pumping and then undergo $\nicefrac{\pi}2$ rotations by a global microwave pulse of frequency 9.2~GHz.
The CZ$_\phi$ gate is then implemented by using Rydberg-excitation lasers acting on both atoms with individual addressing.
The extra phase of~$\phi$ is compensated by introducing a Stark shift to $\ket1$ using the blue-detuned $\sigma_+$ laser.
A final $\nicefrac{\pi}2$ rotation of the target qubit is effected by a combination of the global microwave pulse and the focussed $\sigma_+$ laser.
This sequence of one- and two-qubit operations prepares a Bell state with fidelity $F_\text g\sim0.955$, which is estimated by measuring populations of $\ket{00}$ and $\ket{11}$ and the coherence between them.
This value of fidelity is reported after removing state-preparation-and-measurement (SPAM) errors and errors in single-qubit operations. 
Additionally, this gate implementation reports $T_\text g=0.8~\upmu$s.
This completes our explanation of this state-of-the-art experiment~\cite{GSS+22}.

\subsection{Adiabatic Rydberg-blockade gate in simulation}
\label{subsec:adiabaticgate}
In this subsection, we provide salient background on how a two-qubit neutral-atom gate is achieved by employing adiabatic Rydberg blockade.
First we briefly discuss two proposals for adiabatic population transfer using off-resonant laser pulses acting on a single atom. 
Next we explain how such a population transfer, along with Rydberg-blockade interaction, realizes a CZ$_\phi$ gate in a pair of atoms.
We then present the Hamiltonian whose evolution generates this gate.
Finally, we state the estimated performance of this gate based on numerical simulation.

Two common techniques for adiabatic population transfer in a two-level atom are adiabatic rapid passage~(ARP) and stimulated Raman adiabatic passage~(STIRAP)~\cite{BTE+20}.
The ARP technique, as first studied in the field of nuclear magnetic resonance, involves slowly sweeping the frequency of the electromagnetic field~($\omega$) or the atomic energy separation~($\omega_0$) across resonance~\cite{WS88}.
The condition for this phenomena is~\cite{CF84}
\begin{equation}
\label{eq:ARP_condition}
\nicefrac{\omega_1}{T_2} \ll \frac{\text{d}}{\text{d}t}|\omega-\omega_0| \ll \omega_1^2,
\end{equation}
where $T_2$ is the transverse relaxation time and $\Omega_1$ is the Rabi frequency for the field.
On the other hand, STIRAP is a two-photon excitation process relying on a pair of off-resonant coherent pulses, which are partially overlapping and applied counter-intuitively on the atom~\cite{BTS98}.

Population transfer, via ARP, between~$\ket1$ and~$\ket{\text r}$ is typically executed by a slowly-varying chirped laser pulse applied for a duration~$T$~\cite{MK01}.
This Rydberg-excitation pulse, labelled by $\ell=0$, is associated with time-dependent Rabi frequency~$\Omega\operatorname{e}^{\text i\varphi}$ and detuning~$\Delta$ satisfying the adiabatic condition
\begin{equation}
\label{eq:adiabatic_conditionARP}
\dot{\Theta}\ll\sqrt{\Omega^2+\Delta^2}/2\pi,\;
\Theta:=\operatorname{tan}^{-1}\nicefrac{\Delta}{\Omega},\;
\forall t\in[0,T],
\end{equation}
for~$\Theta$ the mixing angle.
The interaction-picture Hamiltonian ($\hbar\equiv1$) for this process is
\begin{equation}
\frac{H^{\text{d}}(t)}{\hbar}=\frac{\Omega}{2}[\operatorname{e}^{\text i\varphi}\ketbra{1}{\text r}+\operatorname{e}^{-\text i\varphi}\ketbra{\text{r}}{1}] +\frac{\Delta}{2}[\ketbra{\text{r}}{\text{r}}-\ketbra{1}{1}],
\label{eq:Hamiltonian_ARP}
\end{equation}
where $\Omega$ incorporates the slowly-varying envelope of the pulse and $\Delta$ describes its chirping.

In STIRAP, a pair of overlapping laser pulses is used to drive the $\ket1\leftrightarrow\ket{\text{r}}$ transition by virtually employing an intermediate state~$\ket{\text{p}}$~\cite{BTS98}.
One blue-detuned laser pulse driving $\ket1 \leftrightarrow \ket{\text{p}}$ and one red-detuned laser pulse driving $\ket{\text{p}} \leftrightarrow \ket{\text{r}}$, denoted by $\ell=\text{B}$ and $\ell=\text{R}$, respectively, are applied to the atom for a total duration~$T$.  
Associating the laser~$\ell$ with a time-dependent Rabi frequency $\Omega_\ell$ and time-dependent detuning $\Delta_\ell$, this system is described by a Hamiltonian
\begin{align}
H^{\text{q}}(t)
=&\frac12\left(\Omega_\text{B}\ketbra{\text p}{1}+\Omega_\text{R}\ketbra{\text{r}}{\text{p}}\right)+\text{hc}  \nonumber \\
&+\Delta_\text{B}\ketbra{\text{p}}{\text{p}}+\Delta_\text{BR}\ketbra{\text{r}}{\text{r}},
\label{eq:Hamiltonian_STIRAP}
\end{align}
where the two-photon detuning $\Delta_\text{BR} :=\Delta_\text B-\Delta_\text R$.
Using a pair of counterintuitive and partially-overlapping pulses with two-photon resonance, this system adiabatically follows the dark eigenstate of~$H^{\text{q}}(t)$.
This process yields a highly-efficient population inversion between $\ket 1$ and $\ket{\text r}$, with negligible excitation of~$\ket{\text{p}}$.
Additionally, an adiabatic elimination of~$\ket{\text{p}}$, under the condition
\begin{equation}
\label{eq:adiabatic_elimination}
	|\Delta_\text B| \gg |\Omega_\text B|, |\Omega_\text R|,\; 	\forall t\in[0,T],
\end{equation}
yields an effective two-level description for $H^{\text{q}}(t)$~\cite{BPM07}. 

The adiabatic-gate procedure involves simultaneously driving a pair of atoms using a double pulse sequence that returns the atoms to their initial state with conditional phase accumulation~\cite{BSY+13}.
This pulse sequence is symmetric about $t=\nicefrac{T_\text g}{2}$ and constructed by concatenating two adiabatic pulses, with each applied for half the gate time, i.e., $T=\nicefrac{T_\text g}{2}$. 
In the two-qubit computational basis, this procedure yields a controlled-phase gate with the unitary operation 
\begin{align}
\label{eq:CZphiunitary}
    U(T_\text g)=&\ket{00}\!\bra{00}+\mathrm{e}^{\mathrm{i}\phi_{01}}\ket{01}\!\bra{01}+\mathrm{e}^{\mathrm{i}\phi_{10}}\ket{10}\!\bra{10}\nonumber\\
    &+\mathrm{e}^{\mathrm{i}\phi_{11}}\ket{11}\!\bra{11},
\end{align}
where $\phi_{\imath\jmath}$ is the phase accumulated by~$\ket{\imath\jmath}$ over~$T_\text g$.
The phases satisfy $\phi_{01}=\phi_{10}=\pi$ if the shapes of the two pulses in a double-ARP sequence are identical, whereas $\phi_{01}=\phi_{10}=0$ for a double-STIRAP sequence as~$\ket{10}$ and~$\ket{01}$ are dark states with zero eigenenergies~\cite{BTE+20}. 
Intensities and detunings of the lasers are carefully designed to ensure that $\phi_{11}=\pi$, and to consequently execute $\text{CZ}_\pi$ and CZ$_0$ (= CZ) operations~\eqref{eq:CZphimatrix} with double-ARP and double-STIRAP sequences, respectively.
We denote these two gates as ARP $\text{CZ}_\pi$ gate and STIRAP CZ gate.

In simulation, an adiabatic Rydberg-blockade gate is executed by solving the two-atom dynamics over~$T_\text g$.
Typically, the effective contribution of all ground hyperfine levels, besides $\ket0$ and $\ket1$, to the dynamics is modelled as a decay channel to a single ground state~\cite{SBD+20}, which we denote as $\ket{\text g}$.
The two-atom system is mathematically described by the Hamiltonian~\cite{SBD+20}
\begin{equation}
H_\text{B}^{\text e}(t)=H^{\text e}(t)\otimes\mathbb{1}+\mathbb{1}\otimes H^{\text e}(t)+B\ket{\text{rr}}\!\bra{\text{rr}},
\label{eq:Hamiltonian_adiabatic}
\end{equation}
where the label e=d for the one-photon or dipole case~(\ref{eq:Hamiltonian_ARP}) and e=q for the two-photon or quadrupole case~(\ref{eq:Hamiltonian_STIRAP}), with $\varphi=0$ and a constant~$B$ satisfying the Rydberg-blockade condition 
\begin{equation}
\label{eq:blockade_condition}
	B \gg \Omega, \Omega_\text{B}, \Omega_\text{R},~\forall t\in[0,T_\text g].
\end{equation}
This condition ensures that for an initial state~$\ket{11}$, the two-atom system undergoes one cycle of efficient population transfer between $\ket{11}$ and the symmetric state
\begin{equation}
\label{eq:plus_state}
    \ket + :=\frac{\ket{\text{1r}}+\ket{\text{r1}}}{\sqrt{2}},
\end{equation}
with a negligible population in~$\ket{\text{rr}}$, and the anti-symmetric state $\ket -:=\frac{\ket{\text{1r}}-\ket{\text{r1}}}{\sqrt{2}}$ being uncoupled to the rests.

The above gate procedure is predicted to yield robust and high-fidelity adiabatic gates on a pair of neutral alkali atoms~\cite{MMM+14,SBD+20}.
Greedy optimization over pulse parameters predicts $F_\text g\sim0.98$ with ARP pulses~\cite{MMM+14}.
A numerical analysis of this gate procedure, with Cs atomic parameters and decay from excited levels, yields $F_\text g=0.998$ over $T_\text g=1~\upmu$s using globally optimized STIRAP-inspired pulses~\cite{SBD+20}.
Furthermore, this analysis predicts $F_\text g=0.9994$ over $T_\text g=0.54~\upmu$s for ARP pulses.
Although these gates have high fidelities, they are fundamentally slow due to the use of adiabatic pulses.
\begin{table}[]
    \centering
    \begin{tabular}{|c|c|}
         \hline
         ARP& adiabatic rapid passage\\
         \hline
         TQD & transitionless quantum driving\\
         \hline
         STIRAP & stimulated Raman adiabatic passage\\
          \hline
        MC & Monte Carlo\\
        \hline
         cTQD & constrained TQD\\
         \hline
         GKSL & Gorini-Kossakowski-Sudarshan-Lindblad \\
        \hline
        SPAM & state preparation and measurement \\
        \hline
        DE & differential evolution\\
        \hline
        LCG & linearly chirped Gaussian\\
        \hline
        ZCHG & zero-chirp hyper-Gaussian\\
        \hline
    \end{tabular}
    \caption{A list of abbreviations.}
    \label{tab:my_label}
\end{table}

\subsection{Estimating fidelity in simulation}
\label{subsec:characterize}
Here we discuss the typical procedures for simulating the generation of an approximate Bell state and estimating the state fidelity for two cases, namely, with and without technical imperfections and spontaneous emission.
We use the term `imperfections' to refer to fluctuations of laser intensity, atomic position and detunings, which are in turn affected by magnetic-field and temperature variations.
We start by introducing the sequence of operations that prepares an ideal Bell state and then define fidelity of a non-ideal state generated by this sequence.
Next we elaborate on the process of generating the requisite evolution for a two-qubit operation using the two-atom Hamiltonian.
At the end, we discuss how to incorporate parameter fluctuations into the Hamiltonian dynamics.

A Bell state is generated from the ground state~$\ket{11}$ using a sequence of single- and two-qubit gate operations.
The two-atom system is initially prepared in an equal superposition of the computational basis states as
\begin{equation}
\label{eq:initial_state}
	\ket{\psi_0}=\frac{(\ket0-\ket1)\otimes(\ket0-\ket1)}{2}
\end{equation}
by applying Hadamard (H) gates to each of the qubits in~$\ket{11}$, which are denoted as control and target, respectively.
The final state~$\ket{\psi_\text f}$, which ideally is a Bell state~\cite{NC02} 
\begin{equation}
\label{eq:Bell_state}
	\ket{\beta_{\imath\jmath}}
    :=\left(\ket{0,\jmath}
        +(-1)^\imath\ket{1,1-\jmath}\right)/\sqrt{2},
        \;  \forall \imath,\jmath\in\{0,1\},
\end{equation}   
is then prepared from $\ket{\psi_0}$ by first applying a controlled-phase gate and then another H gate to the target qubit.
Particularly, ideal CZ and $\text{CZ}_\pi$ gates yield the Bell states $\ket{\beta_{01}}$ and $\ket{\beta_{00}}$, respectively~\cite{SBD+20}. 

The Bell-state preparation fidelity is defined as the overlap between the final two-atom state~$\ket{\psi_\text f}$ and the corresponding ideal Bell-state~$\ket{\beta_{\imath\jmath}}$.
Under assumptions of ideal and instantaneous H gates and non-ideal controlled-phase gate, which incorporates only the fundamental imperfection due to finite~$B$, the intrinsic fidelity of the given controlled-phase gate is 
\begin{equation}
\label{eq:fidelity_intrinsic}
 F^0_\text g := F^0_{\imath\jmath}
    =|\bra{\beta_{\imath\jmath}}\psi_\text f\rangle|^2,
\end{equation} 
where $F^0_{\imath\jmath}$ is the intrinsic Bell-state preparation fidelity.
Furthermore, the realistic fidelity~$F_\text g$ is measured in terms of the realistic Bell-state preparation fidelity~$F_{\imath\jmath}$ as
\begin{equation}
\label{eq:fidelity_realistic}
F_\text g:= F_{\imath\jmath}
    =\bra{\beta_{\imath\jmath}}\rho_\text f\ket{\beta_{\imath\jmath}},
\end{equation} 
where $\rho_\text f$ denotes the final density matrix of the two-atom system obtained via non-unitary dynamics in the presence of decay and imperfections.
In the following, we explain how realistic gate operations are derived from open-system evolutions.

The evolution of a time-dependent Hamiltonian, which represents the dynamics of a coupled two-atom system, over time~$T_\text g$ yields a two-qubit entangling gate.
For an interaction picture Hamiltonian~$H$, two-atom density matrix~$\rho(t)$ and a `superoperator'~$\mathscr{L}[\rho(t)]$, the dynamics of a Rydberg-blockade gate is simulated by integrating the two-atom Gorini-Kossakowski-Sudarshan-Lindblad (GKSL) equation~\cite{MGE+10,ZGI+12,DBL+18,SBD+20}
\begin{equation}
\label{eq:Lindbladian}
	\frac{\text{d}\rho(t)}{\text dt}=\operatorname{i}\left[\rho(t),H\right]+\mathscr{L}[\rho(t)]\otimes\mathbb{1}+\mathbb{1}\otimes\mathscr{L}[\rho(t)].
\end{equation}
The superoperator corresponding to all decay channels is 
\begin{equation}
	\mathscr{L}[\rho(t)] := \sum_{j,k<j}
	c_{jk}\rho(t) c_{jk}^{\dagger} - \frac12\left\{c_{jk}^{\dagger} c_{jk}, \rho(t) \right\},
\end{equation}
where~$\{,\!\}$ denotes an anticommutator and
the collapse operator 
\begin{equation}
\label{eq:collapseop}
c_{jk}:=\sqrt{b_{jk}\gamma_j}\ketbra{k}{j}
\end{equation}
denotes the radiative decay channel from a higher energy state~$\ket{j}$ to a lower energy state~$\ket{k}$ with a decay rate $b_{jk}\gamma_j$.
The total decay rate from~$\ket{j}$ is $\gamma_j$ and the coefficient~$b_{jk}$, called branching ratio, denotes the fraction of decay to one of the lower energy states. 
Particularly, for each atom, the relevant high-energy states are $\ket{\text{r}}$ and $\ket{\text{p}}$ with decay rates $\gamma_\text r$ and $\gamma_\text p$, respectively, where the latter has a dominant contribution to a two-photon Rydberg excitation process~\cite{LKS+19, GKG+19}.

To account for imperfections in quantum gate operations, Monte-Carlo~(MC) simulation of the GKSL equation~(\ref{eq:Lindbladian}) is performed with realistic parameter fluctuations~\cite{ZGI+12,DBL+18,MGE+10}.
These imperfections include thermal motion of atoms in traps, laser (intensity and phase) fluctuation, Doppler dephasing of Rydberg state, magnetic field fluctuations and dynamic Stark shifts of atomic energy levels.
A detailed analysis of these technical imperfections shows that the dominant gate errors are atomic motion and dephasing, which arise predominantly due to non-zero atomic temperature~\cite{GKG+19}.
Below we describe modelling of fluctuations in the Rabi frequency and detuning of each Rydberg-excitation laser, and finally explain how to estimate fidelity from MC simulation.

The two major sources of Rabi-frequency fluctuations are atomic temperature and laser-intensity fluctuations.
The atomic position~$\bm{r}:=(x,y,z)$ at each trap site varies according to normal distributions as~\cite{ZGI+12}
\begin{equation}
\label{eq:atom_position}
x\sim\mathcal{N}(0,\sigma_x^2),\;
y\sim\mathcal{N}(0,\sigma_y^2),\;
z\sim\mathcal{N}(0,\sigma_z^2),
\end{equation}
where $\sigma_x$, $\sigma_y$ and $\sigma_z$ depend on trap parameters and atomic temperature.
These variations, along with the minimum beam waist~$w_\ell$ of the Rydberg-excitation laser~$\ell$, lead to fluctuations in the spatial form factor~\cite{ZGI+12,GKG+19}
\begin{equation}
\label{eq:space_function}
p_\ell(\bm r)
=\frac{\operatorname{e}^{-\frac{x^2+y^2}{w_\ell^2\left(1+\nicefrac{z^2}{L_\ell^2}\right)}}}{\sqrt{1+\nicefrac{z^2}{L_\ell^2}}}
\end{equation}
of the Rabi frequency, where the corresponding Rayleigh length $L_\ell:=\nicefrac{\pi w_\ell^2}{\lambda_\ell}$.
Additionally, a laser-intensity fluctuations is modelled as the factor~\cite{MGE+10,ZGI+12,DBL+18}
\begin{equation}
\label{eq:intensity_function}
f_\ell \sim\sqrt{1+\mathcal{N}\left(0,(\sigma^\text{i}_\ell)^2\right)},
\end{equation}
with~$\sigma^\text{i}_\ell$ being the standard deviation of intensity for the laser~$\ell$.

The dephasing of the Rydberg state relative to the ground state arises mainly due to the Doppler effect and magnetic-field variation~\cite{SZG+11}; this dephasing process is modelled as fluctuations in laser detuning~\cite{ZGI+12}. 
For atomic mass~$M_\text a$ and Boltzmann constant~$k_\text B$, the dephasing rate due to the Doppler effect is~\cite{ZGI+12, SZG+11, DBL+18} 
\begin{equation}
	\label{eq:T2_Doppler}
	\sigma^\text{D,e}=\sqrt{2}\left(T_2^\text{D,e}\right)^{-1}:=
	\sqrt{2}\left(\frac{\sqrt{\nicefrac{2M_\text{a}}{k_\text{B}T_\text a}}}{k^\text{e}_\text{eff}}\right)^{-1},
\end{equation}
where $k^\text{d}_\text{eff}:=\nicefrac{2\pi}{\lambda_0}$.
In the presence of a magnetic field fluctuating with a standard deviation~$\sigma_\text B$ and a Rydberg state with magnetic quantum number~$m_j$, the dephasing rate~\cite{SZG+11,ZGI+12} 
\begin{equation}
	\label{eq:T2_magnetic}
	\sigma^\text{m,e}=\sqrt{2}\left(T_2^\text{m,e}\right)^{-1}:=
	\sqrt{2}\left(\frac{2^{3/2}\pi\hbar}{|g_\text rm_j-g_1m_f|\upmu_\text B\sigma_\text B}\right)^{-1},
\end{equation}
for $g_1m_f=0$~\eqref{eq:qubit}.
These two decoherence processes are then modelled as random shifts~\cite{ZGI+12,DBL+18}
\begin{equation}
\label{eq:Rydberg_dephasing}
\Delta_\text D\sim\mathcal{N}\left(0,(\sigma^\text{D,e})^2\right),\;
\Delta_\text m\sim\mathcal{N}\left(0,(\sigma^\text{m,e})^2\right),
\end{equation}
of the Rydberg-level detuning for Doppler sensitivity and magnetic sensitivity, respectively.
Other sources for this decoherence include laser phase noise~\cite{DBL+18} and a finite value for~$\gamma_\text r$~\cite{GKG+19}, where the latter is already included in $c_{\text{r}k}$~\eqref{eq:collapseop}.

MC simulation of a quantum gate involves solving the GKSL equation~\eqref{eq:Lindbladian} multiple times using different values for Hamiltonian coefficients, and reporting the average solution~\cite{ZGI+12,DBL+18,MGE+10}.
To account for atomic-position and laser-intensity fluctuations in each MC run, the time-dependent Rabi frequency function for each laser~$\ell$ is multiplied by two values sampled from~$p_\ell(\bm r)$ and~$f_\ell$.
Furthermore, random samples from $\Delta_\text D$ and $\Delta_\text B$ are added to the time-dependent detuning in order to include dephasing effects in every MC run.
The estimated fidelity in the presence of spontaneous emission and these imperfections is the average of fidelities calculated in all MC runs by solving the corresponding GKSL equation and evaluating $F_\text g$~\eqref{eq:fidelity_realistic}.

\subsection{Transitionless quantum driving}
\label{subsec:TQD}
TQD is an alternative to using adiabatic pulses and narrow-band on-resonant pulses for efficient population transfer between atomic energy levels~\cite{DR03, Berry09}.
In this subsection, we summarize relevant background on TQD and its applications to quantum control. 
We begin by introducing the control Hamiltonian and the TQD Hamiltonian for a general quantum system.
Next we provide expressions of the control and TQD Hamiltonians for a two-level system and discuss possible experimental realizations for this TQD Hamiltonian.
Finally, we state examples of quantum-control problems that employ this technique. 

In TQD, the quantum system follows the instantaneous eigenstates of an adiabatic Hamiltonian for all integration times.
Given an adiabatic~$H(t)$ with instantaneous eigenstates~$\{\ket{E(t)}\}$ and corresponding eigenenergies~$\{\ket{\epsilon(t)}\}$, the transitionless dynamics is achieved by adding a TQD control Hamiltonian~\cite{Berry09}
\begin{equation}
\label{eq:Hc}
    H_\text{c}(t)
        =\text{i}\sum_E \left(\ketbra{\dot E(t)}{E(t)}-\braket{E(t)|\dot E(t)}\ketbra{E(t)}{E(t)}\right)
\end{equation}
to~$H(t)$. 
The resultant TQD Hamiltonian
\begin{equation}
\label{eq:HTQD}
	\check{H}(t)=H(t)+ H_\text{c}(t)
\end{equation}
non-adiabatically drives the system along an adiabatic path of $H(t)$ by effectively cancelling the transitions between~$\{\ket{E(t)}\}$. 
This technique, which we refer to as the ``TQD technique", is often used as a quantum control tool for speeding up adiabatic processes. 

The TQD technique has been applied to an ARP Hamiltonian to speed up population inversion between ground and excited levels of a two-level system~\cite{CLR+10}.
To this end, the control Hamiltonian is calculated using two eigenstates.
For the two-level atomic system described by $H^\text{d}(t)$~\eqref{eq:Hamiltonian_ARP},
the eigenstates are
\begin{align}
\label{eq:Estates}
\ket{E_+(t)}&=\operatorname{cos}\nicefrac{\Theta}{2}\ket{1}+\operatorname{e}^{\text i\varphi}\operatorname{sin}\nicefrac{\Theta}{2}\ket{\text{r}},  \nonumber\\
\ket{E_-(t)}&=-\operatorname{sin}\nicefrac{\Theta}{2}\ket{1}+\operatorname{e}^{\text i\varphi}\operatorname{cos}\nicefrac{\Theta}{2}\ket{\text{r}},
\end{align}
and their corresponding eigenenergies are
\begin{equation}
\label{eq:Eenergies}
	\epsilon_{\pm}(t)=\pm\nicefrac{\sqrt{\Delta^2+\Omega^2}}{2}.
\end{equation}
Assuming a time-independent~$\varphi$ and assigning $\Omega_\text{c}=\dot\Theta$, the control Hamiltonian~\eqref{eq:Hc} is then derived as 
\begin{equation}
\label{eq:Hc_ARP}
H_\text{c}^\text{d}(t)=-\text{i}\operatorname{e}^{\text i\varphi}\frac{\Omega_\text{c}}{2}\ket{1}\!\bra{\text{r}}+\text{hc}.
\end{equation}
In a basis spanned by $\{\ket{E_\pm(t)}\}$, only the off-diagonal terms of this Hamiltonian are non-zero, which signifies that the impact of $H_\text{c}^\text{d}(t)$ is to counteract the non-adiabatic couplings of the eigenstates of $H^\text{d}(t)$.

The TQD Hamiltonian for the two-level system is then calculated by adding $H_\text{c}^\text{d}(t)$ to $H^\text{d}(t)$, and is experimentally realized by either using an additional laser~\cite{CLR+10} or modifying the existing laser~\cite{BVM+12}.
Mathematically, the TQD of a two-level system is represented by
\begin{subequations}
\label{eq:HTQD_ARP}
\begin{align}
\check{H}^{\text{d}}(t)=& \frac{\operatorname{e}^{\text i\varphi} }{2}(\Omega -\text{i}\Omega_\text{c})\ketbra{1}{\text r}+\text{hc}+\frac{\Delta}{2}[-\ketbra{1}{1}+\ketbra{\text{r}}{\text{r}}]\\
\xrightarrow{U_\theta(t)}& \frac{\Omega'}{2}[\operatorname{e}^{\text i\varphi}\ketbra{1}{\text r}+\operatorname{e}^{-\text i\varphi}\ketbra{\text{r}}{1}] +\frac{\Delta'}{2}[\ketbra{\text{r}}{\text{r}}-\ketbra{1}{1}],
\end{align}
\end{subequations}
where 
\begin{equation}
    \Omega':=\sqrt{\Omega^2+\Omega_\text{c}^2},\;
    \Delta':=\Delta+\dot\theta,
\end{equation}
for $\theta:=\operatorname{tan}^{-1}\nicefrac{\Omega_\text{c}}{\Omega}$, and 
Eq.~\ref{eq:HTQD_ARP}(b) is derived from Eq.~\ref{eq:HTQD_ARP}(a) by appropriate unitary transformation using $U_\theta(t):=\operatorname{diag}(1, \text{e}^{\text{i}\theta})$.
The implementation of Eq.~\ref{eq:HTQD_ARP}(a) employs two lasers with the same frequency, orthogonal polarization and different time-dependent functions for intensities, whereas Eq.~\ref{eq:HTQD_ARP}(b) is realized by modifying the time-dependent intensity and detuning of the original laser.
A geometric interpretation for TQD of a similar two-level system, i.e., a spin~\nicefrac12 particle in a rotating magnetic field, using trajectories on a Bloch sphere is also provided~\cite{OHNK16}.

Quantum control based on TQD has been used to transfer population between atomic states, create entangled states and design quantum gates in short times~\cite{CWH+16,WJZ17,MKS+18,WZX+19}.
A particular experiment demonstrates a robust population transfer in cold neutral atoms using TQD STIRAP, which is faster than conventional STIRAP~\cite{DLL+16}. 
This work also highlights the speed-up limit of TQD over adiabatic driving and compares the resource requirement for both processes.
Another work proposed a procedure for generating a three-qubit entangled state in Rydberg atoms using TQD and ground-state blockade mechanism~\cite{SLJ+17}. 
Motivated by these theoretical and experimental works, we use the TQD technique to speed up adiabatic Rydberg-blockade gates.
\section{Approach}
\label{sec:approach}
In this section, we discuss our approach for designing a $\text{CZ}_\phi$~\eqref{eq:CZphimatrix} gate using our modified TQD technique and assessing the gate performance.
We begin by explaining our detailed model for the two-atom system, including effects of spontaneous emission and technical imperfections.
Then we mathematically describe our gate procedure.
Finally, we elaborate on our numerical simulation for the two-qubit gate.

\subsection{Model}
\label{subsec:model}
We adapt the model for a pair of laser-driven neutral atoms~\cite{SBD+20}
and make our model comprehensive by incorporating spontaneous emission and  major technical imperfections~\cite{GKG+19,GSS+22}.
Here we first present our model for two trapped atoms
whose positions fluctuate due to non-ideal trapping conditions.
For each of these atoms, we then consider two different physical processes,
namely,
two types of Rydberg-excitation processes,
one for dipole and the other for quadrupole driving.
We use numerical values for the atomic and experimental parameters that are feasible with state-of-the-art instruments, see~\cref{tab:parameters}.
\begin{table*}[]
\begin{center}
\begin{tabular}{|c|c|c|c|c|c|c|c|c|c|}
\hline
 & \multicolumn{4}{|c|}{Laser}
 & \multicolumn{3}{|c|}{Rydberg state}
 & \multicolumn{2}{|c|}{Rydberg blockade}\\
\hline 
Excitation & Wavelength & Wavevector & Waist & Rayleigh length & Atomic level & $g_j$ & Lifetime & $\nicefrac{B}{2\pi}$ & Distance\\
\hline
\hline
Dipole & 
$\lambda_0=319$~nm & $19.7~\upmu\text{m}^{-1}$ & $w_0=2.5~\upmu$m & $L_0=61.5~\upmu$m &
$\ket{112\text{p}_{\nicefrac32},m_j=\nicefrac32}$ & \nicefrac43 & $593~\upmu$s & 
3 GHz & $3.05~\upmu$m \\
\hline
Quadrupole & 
\shortstack{$\lambda_\text B=459$ nm \\ $\lambda_\text R=1038$ nm} & 7.63 $\upmu\text{m}^{-1}$ & \shortstack{$w_\text B=3~\upmu$m \\ $w_\text R=3~\upmu$m} & \shortstack{$L_\text B=61.6~\upmu$m \\ $L_\text R=27.2~\upmu$m} &
$\ket{112\text{d}_{\nicefrac52}, m_j=\nicefrac{5}{2}}$ & \nicefrac65 &$367~\upmu$s & 
2 GHz & $3.4~\upmu$m \\ 
\hline
\end{tabular}
\end{center}
\caption{Atomic and experimental parameters used in our simulations}
\label{tab:parameters}
\end{table*}

\begin{figure}
   	 \includegraphics[width=.45\linewidth]{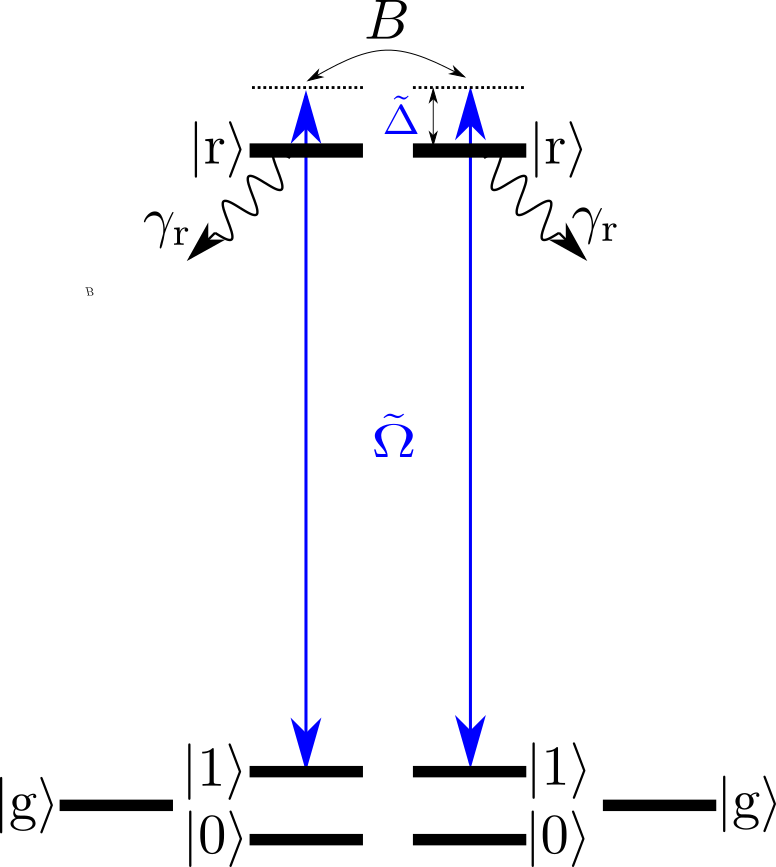}
   	 \includegraphics[width=.45\linewidth]{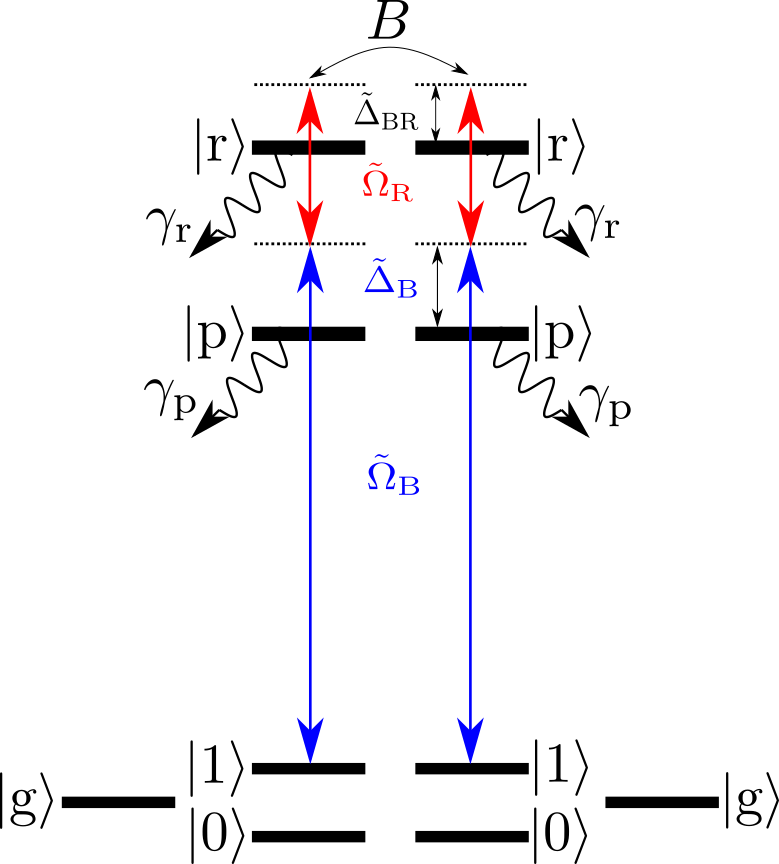}\\
	 (a)~~~~~~~~~~~~~~~~~~~~~~~~~~~~~~~~~(b)
  \caption{A pair of Cs atoms driven by detuned, broadband laser pulses: 
  (a) Each atom is excited to its Rydberg level~$\ket{\text{r}}$ by a dipole-excitation process with time-dependent Rabi frequency~$\tilde\Omega$ and detuning~$\tilde\Delta$. 
  The decay rate~$\gamma_{\text{r}}$ of state~$\ket{\text{r}}$ is $\nicefrac{1}{593}~\upmu$s$^{-1}$ with branching ratios $b_{\text{r}0}=b_{\text{r}1}=\nicefrac{1}{16}$ and $b_{\text{rg}}=\nicefrac{7}{8}$.
The blockade shift between the atoms is 3~GHz.
  (b) Each atom is excited by a quadrupole-excitation process, with time-dependent $\tilde\Omega_\text B$ and $\tilde\Omega_\text R$, via the intermediate state~$\ket{\text{p}}$ and single-photon detuning~$\tilde\Delta_\text B$.
  The two-photon detuning~$\tilde{\Delta}_\text{BR}:=\tilde\Delta_\text B-\tilde\Delta_\text R$.
  The decay rate~$\gamma_\text{r}$ is $\nicefrac{1}{367}~\upmu$s$^{-1}$, and the decay rate~$\gamma_\text{p}$ of state~$\ket{\text{p}}$ is $\nicefrac{1}{0.155}~\upmu$s$^{-1}$ with branching ratios $b_{\text{r}0}=b_{\text{r}1}=\nicefrac1{32}$, $b_{\text{rg}}=\nicefrac{7}{16}$ and $b_{\text{rp}}=\nicefrac12$.
  The blockade shift between the atoms is 2~GHz.
  Other relevant parameters and their values are provided in~\cref{tab:parameters}.
}
   \label{fig:model}
\end{figure}

We consider two cold ($T_\text a=5~\upmu$K~\cite{GSS+22}) Cs atoms trapped in an optical lattice at a separation $\le3.4~\upmu$m and experiencing a strong dipole-dipole interaction of strength $\nicefrac{B}{2\pi}\ge2~$GHz between them.
Each atom is described as a five-level system
corresponding to the qubit basis states $\ket 0$ and $\ket1$~\eqref{eq:qubit}, effective ground state~$\ket{\text g}$, intermediate exited state~$\ket{\text p}$~\eqref{eq:intermedstate}, which is relevant only for quadrupole driving, and a Rydberg state~$\ket{\text r}$~[\cref{fig:model}]. 
We choose the atomic level with principal quantum number 112 as~$\ket{\text r}$ in our simulations, where the exact choice for the orbital quantum number depends on the excitation process~\cite{ZGI+12}.
We neglect decoherence of the ground hyperfine levels, whose $T_1$ and $T^*_2$ are orders of magnitude greater than other relaxation times of the system~(\cref{subsec:experiment}), but consider radiative decay of the excited levels $\ket{\text{p}}$ and~$\ket{\text{r}}$.
The positional uncertainty of each atom at each trap site follows a three-dimensional Gaussian distribution, about the trap center, with standard deviations $\sigma_x=\sigma_y=0.24~\upmu$m and $\sigma_z=0.92~\upmu$m~\cite{GSS+22}. 
Additionally, we assume ideal optical pumping, hyperfine transitions and state detection because errors from SPAM and fast single-qubit gates are typically accounted for when estimating fidelities in experiments~\cite{GKG+19,LKS+19,GSS+22}.

In the dipole-excitation model for a trapped Cs atom,
the $\ket1\leftrightarrow \ket{\text{r}}$ transition of both atoms is driven simultaneously using an off-resonant, broadband laser, which is associated with a time-dependent Rabi frequency $\tilde{\Omega}(t)$ and detuning $\tilde{\Delta}(t)$; see~\cref{fig:model}(a).
Our model can be implemented using an ultraviolet laser of wavelength~$\lambda_0=319~$nm~\cite{HJP+14,MJL+21} simultaneously focused on both atoms by an advanced beam-scanning mechanism~\cite{GSS+22}.
In order to safely neglect crosstalk errors, we use a beam waist of $w_0=2.5~\upmu$m and interatomic distance of $3.05~\upmu$m in our simulations.
We consider a $5\%$ fluctuation of the laser intensity, i.e.\ $\sigma^\text{i}_0=0.05~\eqref{eq:intensity_function}$, but neglect phase noise because of the current experimental improvements in resonant filtering~\cite{LKO+18}.
A strong Rydberg blockade with $\nicefrac{B}{2\pi}=3$~GHz is achieved between the atom pair by our choice of \begin{equation}
\label{eq:r}
\ket{\text{r}}=\ket{112\text{p}_{\nicefrac32},m_j=\nicefrac32}
\end{equation}
with a decay rate~\cite{ZGI+12} (branching ratios)
\begin{equation}
 \label{eq:decayrates_dip}
    \nicefrac{\gamma_\text r}{2\pi}=\nicefrac{1}{593}~\text{MHz}\;\left(b_{\text{r}0}=b_{\text{r}1}=\nicefrac1{16},\,
b_{\text{rg}}=\nicefrac{7}{8}\right).
\end{equation}
Additionally, we simulate dephasing of $\ket{\text{r}}$ with respect to $\ket1$ using $T_2^\text{D,d}=4~\upmu$s~\eqref{eq:T2_Doppler} and $T_2^\text{m,d}=50~\upmu$s~\eqref{eq:T2_magnetic}, for $\sigma_\text B=10^{-6}$~T~\cite{GKG+19}.

In the quadrupole-transition model, both Cs atoms are simultaneously driven between their ground and Rydberg states using a pair of partially-overlapping laser pulses~\cite{SBD+20}.
This model has been realised using narrowband pulses applied to a 2D lattice~\cite{GKG+19,GSS+22}.
We describe the effect of the blue(red)-detuned laser, having $\lambda_{\text B(\text R)}=459(1038)~$nm, on the $\ket1(\ket{\text p})\leftrightarrow \ket{\text{p}}(\ket{\text{r}})$ transition by time-dependent Rabi frequency $\tilde{\Omega}_{\text B(\text R)}(t)$ and detuning $\tilde{\Delta}_{\text B(\text R)}(t)$; see Fig.~\ref{fig:model}(b).
Each of these two Rydberg-excitation lasers is broadband with a finite beam waist of $w_{\text B(\text R)}=3~\upmu$m.
By choosing an interatomic distance of $3.4~\upmu$m and 
\begin{equation}
\ket{\text{r}}=\ket{112\text{d}_{\nicefrac52},m_j=\nicefrac52},
\end{equation}
we simulate a strong blockade with $\nicefrac{B}{2\pi}=2$~GHz.
We consider spontaneous emission from the two excited levels $\ket{\text{p}}$ and $\ket{\text{r}}$, with decay rates (branching ratios)~\cite{ZGI+12}
\begin{align}
 \label{eq:decayrates_quad}
\nicefrac{\gamma_\text p}{2\pi}&=\nicefrac{1}{\tau_{7 \text{p}_{\nicefrac12}}}\;(b_{\text{p}0}
=b_{\text{p}1}=\nicefrac1{16},\, b_{\text{pg}}=\nicefrac78), \nonumber\\
\nicefrac{\gamma_\text r}{2\pi}&=\nicefrac{1}{367}~\text{MHz}\;(b_{\text{r}0}=b_{\text{r}1}=\nicefrac1{32},\,
b_{\text{rg}}=\nicefrac{7}{16},\,
b_{\text{rp}}=\nicefrac12),  
\end{align}
respectively.
Additional imperfections arising from laser-intensity fluctuations, Doppler effect and magnetic field fluctuations are modelled using parameters $\sigma^\text{i}_\text{B,R}=0.01$~\cite{MLX+15}, $T_2^\text{D,q}=10.5~\upmu$s and $T_2^\text{m,q}=34~\upmu$s, respectively.

\subsection{Mathematics}
\label{subsec:math}
In this subsection, we explain the mathematics behind designing and characterizing a $\text{CZ}_\phi$ gate~\eqref{eq:CZphimatrix}.
We begin by providing a valid Hamiltonian representation for our laser-driven two-atom model~(\S\ref{subsec:model}) and then discuss how we decompose the Hamiltonian into a direct sum.
Next we elaborate our procedure to derive time-dependent coefficients for this Hamiltonian, 
such that the resultant unitary dynamics yields a ground-Rydberg entangled state for an integration time that is faster than the adiabatic time scale.
As this dynamics is only necessary but not sufficient to achieve a high-fidelity, fast $\text{CZ}_\phi$ operation, we propose a recipe to design each time-dependent laser beam as a sequence of phase-shifted pulses.
Finally, we introduce our metric to evaluate a $\text{CZ}_\phi$ gate.

For each of the two Rydberg-excitation models, i.e.\ dipole and quadrupole, we generalize the respective single-atom Hamiltonians~$H^{\text e}(t)$ (\ref{eq:Hamiltonian_ARP},\ref{eq:Hamiltonian_STIRAP}) to account for imperfections in the $\ket 1\leftrightarrow\ket{\text r}$ transition.
To this end, we assume non-adiabatic parametrization for the time-dependent coefficients and include additional coefficients describing parameter fluctuations.
The dipole form of this space-time-dependent Hamiltonian~$\tilde{H}^{\text e}(t,\bm r)$ is
\begin{align}
\label{eq:Hamiltonian_dipole}
\tilde{H}^{\text{d}}(t,\bm r)=&\frac12p_0(\bm r)f_0\tilde{\Omega}(t)\ketbra{1}{\text r} +\text{hc}\nonumber\\
&+\frac{\tilde\Delta(t)+\Delta_\text m +\Delta_\text D}{2}[\ketbra{\text{r}}{\text{r}}-\ketbra{1}{1}],
\end{align}
where fluctuations in Rabi frequency and detuning are described by $p_0(\bm r)f_0$~(\ref{eq:space_function},\ref{eq:intensity_function}) and random shifts $\Delta_\text m +\Delta_\text D$~\eqref{eq:Rydberg_dephasing}, respectively.
Similarly, for the quadrupole transition,
\begin{align}
\label{eq:Hamiltonian_quadrupole}
\tilde H^{\text{q}}(t,\bm r)
=&\frac12p_\text B(\bm r)f_\text B\tilde\Omega_\text{B}(t)\ketbra{\text{p}}{1} +\frac12p_\text R(\bm r)f_\text R\tilde\Omega_\text{R}(t)\ketbra{\text{r}}{\text{p}}  \nonumber \\
&+\text{hc} +\tilde\Delta_\text{B}(t)\ketbra{\text{p}}{\text{p}} \nonumber \\
&+\left(\tilde\Delta_\text{BR}(t)+\Delta_\text m +\Delta_\text D\right)\ketbra{\text{r}}{\text{r}},
\end{align} 
where we assume $\tilde\Delta_\text{BR}\approx0$ and a high~$\tilde\Delta_\text B$ satisfying the adiabatic elimination condition~\eqref{eq:adiabatic_elimination}.
\begin{table*}[ht!]
    \centering
    \begin{tabular}{|c|c|}
    \hline
       $H^\text e (t)$  &  single-atom time-dependent Hamiltonian with LCG or ZCHG pulses\\
       \hline
        $H^\text{e}_\text{B} (t)$  & two-atom time-dependent Hamiltonian with LCG or ZCHG pulses \\
        \hline
        $H^\text{e}_\text{B,eff} (t)$  & effective Hamiltonian obtained by projecting $H^\text{e}_\text{B} (t)$ on to~$\mathscr H_\text{rr}$ \\
        \hline
        $\check H^\text{e}_\text{B,eff} (t)$  & TQD Hamiltonian obtained from $H^\text{e}_\text{B,eff} (t)$ \\
        \hline
        $\tilde H^\text e(t,\bm r)$  &  single-atom space-time-dependent cTQD Hamiltonian with TQD pulses\\
       \hline
        $\tilde H^\text{e}_\text{B} (t,\bm r)$  & two-atom space-time-dependent cTQD Hamiltonian with TQD pulses \\
        \hline
        $\tilde H^\text{e}_\text{B,eff} (t,\bm r)$  & effective Hamiltonian obtained by projecting $\tilde H^\text{e}_\text{B} (t, \bm r)$ on to $\mathscr H_\text{rr}$\\
        \hline
        $H^\text{e}_\text{B} (t,\bm r)$  & two-atom Hamiltonian obtained by introducing space dependence in $H^\text{e}_\text{B} (t)$ \\
        \hline
    \end{tabular}
    \caption{List of relevant Hamiltonians.}
    \label{tab:Hamiltonians}
\end{table*}

Our model for a pair of interacting atoms is thus mathematically represented by a space-time-dependent Hamiltonian, which is valid for our gate procedure under the following three conditions.
This two-atom Hamiltonian 
\begin{equation}
\label{eq:Hamiltonian_diabatic}
\tilde{H}_\text{B}^{\text e}(t, \bm r)=\tilde{H}^{\text e}(t,\bm r)\otimes\mathbb{1}+\mathbb{1}\otimes \tilde{H}^{\text e}(t,\bm r)+B\ket{\text{rr}}\!\bra{\text{rr}},
\end{equation}
whose time-dependent coefficients follow the Rydberg-blockade condition~\eqref{eq:blockade_condition}, acts on a 25-dimensional Hilbert space~$\mathscr{H}$.
To describe a valid CZ$_\phi$ operation over time~$T_\text g$, our gate procedure requires $\tilde{H}_\text{B}^{\text e}(t, \bm r)$ to satisfy the following conditions.
\begin{enumerate}
    \item [C1.] Unitary evolution generated by $\tilde{H}_\text{B}^{\text e}(t,\bm r)$, in the absence of parameter fluctuations, yields an efficient population transfer back to the same initial state~$\ket{11}$ over any $T_\text{g}$.
    \item [C2.] Unitary dynamics for an initial state~$\ket{10}$ also yields an efficient population transfer back to the same initial state over any $T_\text{g}$.
    \item [C3.] Final accumulated phases of initial states belonging to the two-qubit computational basis~\eqref{eq:CZphiunitary} must satisfy the relation
\begin{equation}
 \label{eq:CZphirelation}
	\phi_{11}-2\phi_{10} = (2n+1)\pi,~\forall n\in\mathbb{Z}.
\end{equation}
\end{enumerate}
These conditions together ensure the implementation of a high-fidelity CZ$_\phi$ gate.

We can decompose this 25-dimensional~$\tilde{H}_\text{B}^{\text e}(t, \bm r)$ into a direct sum of one- and two-dimensional Hamiltonians; see Appendix~\ref{app:math} for details.
This is achieved by first identifying the 9-dimensional subspace~$\mathscr{H}_\text{CZ}$, spanned by $\{\ket0,\ket1,\ket{\text r}\}$ of both atoms and decoupled from their other states, which captures the physics of a CZ$_\phi$ gate.
This 9-dimensional subspace further decomposes as 
\begin{align}
9=&1\oplus2\oplus2\oplus4 \approx  1\oplus2\oplus2\oplus2\oplus1\oplus1 \nonumber\\
= & \underbrace{\operatorname{span}\{\ket{00}\}}_{\mathscr{H}_0} \oplus
\underbrace{\operatorname{span}\{\ket{01},\ket{0\text{r}}\}}_{\mathscr{H}_{0\text{r}}} \oplus
\underbrace{\operatorname{span}\{\ket{10},\ket{\text{r}0}\}}_{\mathscr{H}_{\text{r}0}} \nonumber \\
&\oplus\underbrace{\operatorname{span}
	\{\ket{11}, \ket{+}\}}_{\mathscr H_\text{rr}} 
\oplus \operatorname{span}\{\ket{-}\} \oplus \operatorname{span}\{\ket{\text{rr}}\}.
\label{eq:qubitspan}
\end{align}
Each of the two-dimensional Hamiltonians acting on $\mathscr H_\text{0r}$ and $\mathscr H_\text{r0}$ resembles the single-atom Hamiltonian for dipole driving~\eqref{eq:Hamiltonian_ARP}, whereas the two-dimensional Hamiltonian acting on $\mathscr H_\text{rr}$, denoted by $\tilde{H}_\text{B,eff}^\text{e}(t, \bm r)$, generates atom-atom entanglement.
We formalise the mapping from~$\tilde{H}_\text{B}^\text{e}(t, \bm r)$ to~$\tilde{H}_\text{B,eff}^\text{e}(t, \bm r)$ by a `conjugating-channel' transformation as
\begin{equation}
\label{eq:reduction}
\tilde{H}_\text{B,eff}^\text{e}(t, \bm r) \oplus \mathbb0_{23}
=\mathcal{C}_R\left(\tilde{H}_\text{B}^\text{e}(t, \bm r)\right),
\end{equation}
where the conjugating channel
\begin{equation}
\label{eq:conjchannel}
\mathcal{C}_R(\bullet)
:=R\bullet R^\top,
\end{equation}
and the real-valued matrix~$R$ is a composition of projection and permutation operators.

To make TQD feasible for 25-dimensional~$\mathscr{H}$, we modify the standard TQD technique~(\S\ref{subsec:TQD}).
Our modification involves the following:
first derive a TQD Hamiltonian for a low-dimensional subspace of $\mathscr{H}$
and then map this Hamiltonian back to a full 25-dimensional matrix.
This procedure yields time-dependent coefficients for $\tilde{H}_\text{B}^\text{e}(t, \bm r)$ as functions of the coefficients for the adiabatic Hamiltonian $H_\text{B}^\text{e}(t)$~\eqref{eq:Hamiltonian_adiabatic}.

We apply the TQD technique to an adiabatic Hamiltonian acting on a low-dimensional, non-trivial subspace of $\mathscr{H}$.
Here we pick $\mathscr H_\text{rr}$ because Hamiltonian dynamics on this subspace is entangling.
The effective Hamiltonian~$H_\text{B,eff}^\text{e}(t)$ acting on $\mathscr H_\text{rr}$ is calculated as
\begin{equation}
\label{eq:hamil_adia_eff_der}
	H_\text{B,eff}^\text{e}(t) \oplus \mathbb0_{23}
	=\mathcal{C}_R\left(H_\text{B}^\text{e}(t)\right),
\end{equation}
where $\mathcal{C}_R$~\eqref{eq:conjchannel} is justified because the graph representations of $H_\text{B}^\text{e}(t)$ and $\tilde{H}_\text{B}^\text{e}(t, \bm r)$ are isomorphic to each other.
The time-dependent coefficients of~$\tilde{H}_\text{B,eff}^\text{e}(t, \bm r)$, in the absence of parameter fluctuations, are then assigned functions by 
\begin{equation}
\label{eq:checktotilde}
\tilde{H}_\text{B,eff}^\text{e}(t, \bm0)\gets\check{H}_\text{B,eff}^\text{e}(t),
\end{equation}
where $\check{H}_\text{B,eff}^{\text{e}}(t)$ is the TQD Hamiltonian obtained from the adiabatic $H_\text{B,eff}^\text{e}(t)$.

We now utilize this two-dimensional $\tilde{H}_\text{B,eff}^{\text{e}}(t,\bm r)$ to derive expressions for the time-dependent coefficients of the 25-dimensional~$\tilde{H}^\text{e}_\text{B}(t, \bm r)$.
This derivation is achieved by the inverse operation of $\mathcal{C}_R$~\eqref{eq:reduction} as
\begin{align}
\label{eq:consTQDH}
\tilde{H}_\text{B}^\text{e}(t, \bm r)
 =&\mathcal{C}_R^{-1}\left(\tilde{H}_\text{B,eff}^\text{e}(t, \bm r) \oplus \mathbb0_{23}\right).
\end{align}
We refer to this constructed Hamiltonian as a ``constrained-TQD Hamiltonian", where ``constrained-TQD" (cTQD) signifies that the TQD technique is applied to a Hamiltonian acting on only one subspace of the full Hilbert space. 
Due to this unique construction, the unitary dynamics of~$\tilde H_\text{B}^\text{e}(t, \bm r)$, without parameter fluctuations, yields efficient population transfer only between $\ket{11}$ and $\ket{+}$ over any arbitrarily low integration time, thus satisfying C1.
\cref{tab:Hamiltonians} summarizes the relevant Hamiltonians used in this paper.

In order to derive explicit time-dependent functions for $\tilde H_\text{B}^\text{e}(t, \bm r)$, we use popular functions for expressing $H_\text{B}^\text{e}(t)$.
Particularly, for describing ARP~\eqref{eq:Hamiltonian_ARP}, we use electric field of the form of a linearly chirped Gaussian~(LCG) function~\cite{BSY+13,BTE+20}, with
\begin{align}
\label{eq:pulses_LCG}
\Omega(t)=\Omega_0\operatorname{e}^{-\frac{(t-\nicefrac{T}{2})^2}{\tau^2}},\;
\Delta(t)=2 \nicefrac{\Delta_0}{T}(t-\nicefrac{T}{2}),
\end{align}
where the Rabi frequency is Gaussian with peak value~$\Omega_0$, mean $\nicefrac{T}{2}$ and width~$\tau$, and the detuning is linear with peak value~$\Delta_0$.
Furthermore, we describe STIRAP pulses~\eqref{eq:Hamiltonian_STIRAP} by Rabi frequencies with Hyper-Gaussian shapes~\cite{SBD+20} as
\begin{align}
\label{eq:pulses_ZCHG}
\Omega_\text B(t)&=\Omega_\text{B0}\operatorname{e}^{-(t-2\nicefrac{T}{3})^4/\tau^4_\text B},\nonumber\\
\Omega_\text R(t)&=\Omega_\text{R0}\operatorname{e}^{-(t-\nicefrac{T}{3})^4/\tau^4_\text R},
\end{align}
and a constant detuning~$\Delta_\text B(\equiv-\Delta_\text R)$, henceforth referred as zero-chirped hyper-Gaussian~(ZCHG).
The parameters of these functions are judiciously chosen to deliver feasible CZ$_\phi$ gates, as detailed in~\S\ref{subsec:method}.

A CZ$_\phi$ gate implemented by evolving the cTQD Hamiltonian~\eqref{eq:consTQDH} does not guarantee high fidelity over low gate time; this deficiency motivates us to re-design pulse shapes such that C2 and C3 are satisfied as well. 
To do this, we start with a $H_\text{B}^\text{e}(t)$ whose coefficients are single-adiabatic pulse functions~(\ref{eq:pulses_LCG},\ref{eq:pulses_ZCHG}), as opposed to double-adiabatic pulses used for executing CZ and CZ$_\pi$ operations~(\S\ref{subsec:adiabaticgate}).
Thus the time-dependent functions $\tilde\Omega$ and $\tilde\Delta$ ($\tilde\Omega_\text B$, $\tilde\Delta_\text B$, $\tilde\Omega_\text R$ and $\tilde\Delta_\text R$) are expressed in terms of LCG (ZCHG) pulse functions.
We refer to these particular pulses as ``TQD pulses" and represent their functions by~$\tilde\Omega^+(t)$ and $\tilde\Delta^+(t)$ for the dipole case, and by $\tilde\Omega^+_\text B(t)$, $\tilde\Delta^+_\text B(t)$, $\tilde\Omega^+_\text R(t)$ and $\tilde\Delta^+_\text R(t)$ for the quadrupole case.

For implementing a high-fidelity cTQD CZ$_\phi$ gate over any gate time $T_\text g$, we design the time-dependent coefficients of~$\tilde{H}_\text{B}^\text{e}(t, \bm r)$~\eqref{eq:Hamiltonian_diabatic} as sequences of four TQD pulse functions.
To construct this sequence, we modify a gate procedure~\cite{LKS+19} by using TQD pulses, which are broadband, instead of narrowband $2\pi$ pulses.
Specifically, we propose dividing a sequence of TQD pulses into two identical sub-sequences having a relative phase shift of $\phi_\text{R}$ between them.
For each sub-sequence, we employ a pair of TQD pulses, with a time-symmetric envelope and a relative phase shift~$\phi_\text{r}$, instead of restricting to a $2\pi$ pulse~\cite{LKS+19}.
The pulse-pair is designed such that the initial state~$\ket{11}$ returns efficiently back to itself over an integration time of $\nicefrac{T_\text g}{2}$, having accumulated a phase of~$\nicefrac{\phi_{11}}2$.
Contrariwise, for the other two non-trivial initial states $\ket{01}$ and $\ket{10}$, this sub-sequence alone does not ensure efficient population transfer back to the corresponding states after $\nicefrac{T_\text g}{2}$.
By judicious choice of the parameter~$\phi_\text R$, the full sequence efficiently returns the atoms to their corresponding initial states $\ket{01}$ and $\ket{10}$ over time~$T_\text g$ with an accumulated phase~$\phi =\phi_{10} (=\phi_{01})$, thus satisfying C2.
Additionally, our choices for $\phi_\text r$ and $\phi_\text R$ ensure that~$\phi_{10}$ and $\phi_{11}$ satisfy the phase relation~\eqref{eq:CZphirelation} for C3.

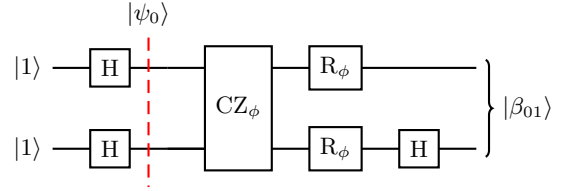
\begin{figure}
\centering
\begin{quantikz}[transparent]
&\lstick{$|{1}\rangle$} & \gate{\text{H}}
 \slice{$|{\psi_0}\rangle$}
 &\qw &\gate[wires=2]{\text{CZ}_\phi} &\gate{\text{R}_\phi} &\qw &\qw 
 \rstick[wires=2]{$|{\beta_{01}}\rangle$} 
\\
&\lstick{$|{1}\rangle$} & \gate{\text{H}}& \qw & \qw & \gate{\text{R}_\phi} & \gate{\text{H}} &\qw 
\end{quantikz}
\caption{A two-qubit quantum circuit for preparing the Bell state~$\beta_{01}$: 
Both qubits are initialized to state~$\ket1$, with upper and lower qubits acting as control and target, respectively.
Each of these qubits undergo H transformation, which ideally yields $\ket{\psi_0}$. 
This state is then subjected to a CZ$_\phi$ gate~\eqref{eq:CZphimatrix}, followed by two R$_\phi$(:=diag(1,e$^{-\text{i}\phi}$)) gates on control and target.
A final H on target ideally yields $\beta_{01}$~\eqref{eq:Bell_state}.
This circuit is a slight modification of the one explained in~\S\ref{subsec:characterize}. 
}
\label{fig:circuit}
\end{figure}

We assess a cTQD CZ$_\phi$ gate by estimating the Bell-state preparation fidelity of a quantum state obtained by applying single-qubit gates and this two-qubit gate on~$\ket{11}$; see~\cref{fig:circuit}.
In each MC run~(\S\ref{subsec:characterize}), we first construct a random $\tilde{H}_\text{B}^\text{e}(t, \bm r)$ by sampling its coefficients from their respective distributions~(\ref{eq:space_function},\ref{eq:intensity_function},\ref{eq:Rydberg_dephasing}) and then integrate the GKSL equation~\eqref{eq:Lindbladian}, with initial state~$\ket{\psi_0}$~\eqref{eq:initial_state} and Cs decay parameters~(\ref{eq:decayrates_dip},\ref{eq:decayrates_quad}), over $T_\text g$ to obtain $\rho(T_\text g)$. 

After executing ideal and instantaneous single-qubit rotations on $\rho(T_\text g)$, the final state
\begin{equation}
\label{eq:final_state}
	\rho_\text f = (\mathbb{1}\otimes\text H)(\text R_\phi \otimes \text R_\phi).\rho(T_\text g)(\text R_\phi^\dagger \otimes \text R_\phi^\dagger)(\mathbb{1}\otimes\text H^\dagger)
\end{equation}
is used to calculate an estimate of the realistic fidelity~\cite{SKK+00}
\begin{equation}
\label{eq:fidelity_gate}
F_\text g := \bra{\beta_{01}}\rho_\text f\ket{\beta_{01}} =\frac{\rho_{0101}+\rho_{1010}}{2}+|\rho_{1001}|,
\end{equation}
where the matrix elements are
\begin{equation}
\label{eq:matrixelements}
\rho_{i_1i_2i_3i_4}=\bra{i_1i_2}\rho_\text f\ket{i_3i_4},\;
i_1,i_2,i_3,i_4\in\{0,1\}.
\end{equation}
We perform 100 MC runs to generate a distribution of fidelity estimates and use the mean to estimate the gate infidelity $1-F_\text g$.
Additionally, we calculate the standard error $\sigma^-_F$ from this distribution, and use $\sigma^-_F$ to represent robustness of the two-qubit gate implementation against spontaneous emission and technical imperfections.


\subsection{Methods}
\label{subsec:method}
In this subsection, we elaborate on the numerical techniques developed and used for our simulation of the interacting two-atom system.
First we present methods to numerically estimate fidelity and time of a CZ$_\phi$ gate~\eqref{eq:CZphimatrix}.
Then we describe our strategy for designing a feasible CZ$_\phi$ gate based on the adiabatic-gate procedure~(\S\ref{subsec:adiabaticgate}), and incorporating imperfections in this gate simulation~(\cref{subsec:characterize}).
Next we provide a numerical recipe that re-defines ``efficiency" in the TQD technique~(\cref{subsec:TQD}). 
Finally, we provide a method to compare adiabatic and cTQD CZ$_\phi$ gates in terms of their resource requirements.

We numerically evaluate the gate fidelity~$F_\text g$ and gate time~$T_\text g$, and use these two quantities to assess the performance of a CZ$_\phi$ gate.
$F_\text g$ is estimated using~\cref{eq:fidelity_gate}, where $\rho(T_\text g)$~\eqref{eq:final_state} is obtained by numerically integrating the GKSL equation~\eqref{eq:Lindbladian} using the Python library~\texttt{QuTiP}.
$T_\text g$ is the integration time and is related to the duration of a single pulse as $T_\text g = 4T$ for a cTQD CZ$_\phi$ gate, as compared to $T_\text g = 2T$ for an adiabatic CZ$_\phi$ gate.
In addition to $F_\text g$ and $T_\text g$, we analyse the trend in fidelity by estimating $F_\text g$ for changing $T_\text g$ values.

We prescribe simple numerical techniques to check feasibility and determine adiabaticity of a Rydberg-blockade CZ$_\phi$ gate.
Our criterion for feasibility dictates that the intrinsic value of gate fidelity~\eqref{eq:fidelity_gate} should satisfy the condition
\begin{equation}
\label{eq:feasibility_condition}
F^0_\text g>\eta^\text e,
\end{equation}
where threshold $\eta^\text{d~(q)}=0.9989~(.989)$.
In the quadrupole driving model, we lower the threshold by 1\% because fidelity exceeding 0.99 requires numerically-optimized pulses~\cite{SBD+20} rather than analytical pulses, which we use here for simplicity.
For a two-level system initiating as an eigenstate~$\ket{E(t)}$~\eqref{eq:Estates} and evolving unitarily, we call the dynamics adiabatic if an instantaneous state~$\ket{\mathcal{I}(t)}$ follows
\begin{equation}
\label{eq:adiabaticity_criteria}
|\braket{E(t)|\mathcal{I}(t)}|^2\ge0.99,\;
\forall t\in[0,T_\text g].
\end{equation}
Based on this condition, the CZ$_\phi$ gate is adiabatic if the dynamics within each two-dimensional subspace~\eqref{eq:qubitspan} is adiabatic.

In order to calculate values for pulse parameters that yields a feasible, adiabatic CZ$_\phi$ gate, we search over specified domains of these parameters.
For LCG pulses~\eqref{eq:pulses_LCG}, we choose 
\begin{align}
\label{eq:parameters_LCG}
&0.1~\upmu\text{s}\le T\le0.25~\upmu\text{s}, \; 10~\text{MHz}\le\nicefrac{\Omega_0}{2\pi}\le25~\text{MHz},\nonumber\\
&20~\text{MHz}\le\nicefrac{\Delta_0}{2\pi}\le50~\text{MHz},\;
0.2T\le\tau\le 0.3 T,
\end{align}
and for ZCHG pulses~\eqref{eq:pulses_ZCHG}, we consider
\begin{align}
\label{eq:parameters_ZCHG}
&50~\text{MHz}\le\nicefrac{\Omega_{\text B0}}{2\pi},\nicefrac{\Omega_{\text R0}}{2\pi}\le300~\text{MHz},\nonumber\\
&100~\text{MHz}\le\nicefrac{\Delta_{\text B}}{2\pi}\le 3000~\text{MHz},\nonumber\\
&0.1~\upmu\text{s}\le T\le5~\upmu\text{s}, \; 0.25T\le\tau_\text B,\tau_\text R \le 0.35T.
\end{align}
For each parameter tuple of LCG (ZCHG) pulse, we simulate the ARP CZ$_\pi$ (STIRAP CZ) gate by unitary evolution of~$H_\text{B}^\text{d}(t)$ $\left(H_\text{B}^\text{q}(t)\right)$~\eqref{eq:Hamiltonian_adiabatic}, with double-LCG (double-ZCHG) pulse functions as its coefficients, and calculate~$F^0_\text g$.
Using an in-house implementation of a global optimization algorithm called differential evolution~(DE)~\cite{SP97}, we then search over the parameter domains to find a tuple satisfying the feasibility condition~\eqref{eq:feasibility_condition}.

We now provide a method for incorporating technical imperfections in our simulation of the adiabatic gates, and then define a non-adiabatic version of this adiabatic-gate procedure.
To account for technical imperfections, we modify the Hamiltonian coefficients by multiplying $p_\ell(\bm r)$~\eqref{eq:space_function} and $f_\ell$~\eqref{eq:intensity_function} to the Rabi frequencies, and adding $\Delta_\text D$ and $\Delta_\text m$~\eqref{eq:Rydberg_dephasing} to the detuning terms.
By introducing space dependence in the Hamiltonian~$H_\text{B}^\text{e}(t)$, we now denote it by~$H_\text{B}^\text{e}(t, \bm r)$.
In the non-adiabatic regime, the dynamics of $H_\text{B}^\text{d}(t, \bm r)$ ($H_\text{B}^\text{q}(t, \bm r)$) yields a non-adiabatic CZ$_\pi$ (CZ) gate, which we denote as a LCG CZ$_\pi$ (ZCHG CZ) gate.

For a transitionlessly driven two-level quantum system, the dynamics of $\check{H}^\text d(t)$~\eqref{eq:HTQD_ARP} yields an efficient population transfer from an initial state~$\ket1$ to the final state~$\ket{\text r}$ in any short time, given that there is no limit on the maximum laser intensity.
To make $\check{H}^\text d(t)$ practical, we re-define an ``efficient" population transfer by imposing that the population of $\ket{\text r}$ at the end of the dynamics exceeds 0.99 and the maximum value for $\Omega'(t)$~(\ref{eq:HTQD_ARP}(b)) is within $10\%$ of the maximum value for $\Omega(t)$~\eqref{eq:Hamiltonian_ARP}.
This upper bound on $\Omega'(t)$ yields a lower limit on the pulse duration, which we denote as~$T_\text{min}$.

We investigate the interplay between two resources, namely maximum (real-valued) Rabi frequency~$\Omega_\text{max}$ and gate time~$T_\text g$, required for achieving high gate fidelity.
To do this, we hypothesize functions relating these resources and empirically estimate the parameters of these functions.
These parameters are then used for comparing different gate procedures.
As a highly-detuned quadrupole driving is mathematically equivalent to a two-level system, we only compare between gates implemented using dipole driving.

By numerically estimating the parameters of a power function
(exponent, coefficient and additive constant)
that relates $\Omega_\text{max}$ and $T_\text g$ for a target fidelity, we determine the scaling of $\Omega_\text{max}$ with decreasing $T_\text g$.
The value of $\Omega_\text{max}$ is equal to $\Omega_0$~\eqref{eq:pulses_LCG} for the ARP CZ$_\pi$ gate, whereas $\Omega_\text{max}$ is a function of $\Omega_0$, $\Delta_0$ and $T_\text g$ for the cTQD CZ$_\phi$ gate and is numerically estimated in our simulation.
For each $T_\text g$, we fix values for other pulse parameters and search for the minimum $\Omega_\text{max}$ required for achieving an ARP CZ$_\pi$ (cTQD CZ$_\phi$) with $F^0_\text g>0.989$ ($F^0_\text g>0.9989$).
In terms of a unit-less gate time~$\bar{T}_\text g:=T_\text g~\upmu\text{s}^{-1}$, power parameter~$p$ and constant frequencies~$\nu_1$ and $\nu_2$, we hypothesize the relation between $\Omega_\text{max}$ and $T_\text g$ as a power function
\begin{equation}
\label{eq:scaling}
	\Omega_\text{max}=\nu_1\bar T^{-p}_\text g+\nu_2,
\end{equation}
in concordance with previous results~\cite{DLL+16}.
The measure of goodness-of-fit for our simulation data to this power function is calculated using the $R^2$ score.
We quantify and compare the performance of the two gates, namely ARP CZ$_\pi$ and cTQD CZ$_\phi$, based on the estimate of~$p$ for each case.

We evaluate the speedup of a cTQD CZ$_\phi$ gate over an ARP CZ$_\pi$ gate, both yielding $F^0_\text g>0.99$, for increasing values of~$\Omega_\text{max}$.
Keeping~$\Omega_\text{max}$ almost same (within 10\%) for both of these gates, we calculate speedup as the ratio of $T_\text g$s for ARP and cTQD gates.
Based on previous results~\cite{DLL+16}, we expect the speedup value to saturate for both high and low magnitudes of $\Omega_\text{max}$.
Thus we can hypothesise the relation between a unit-less maximum Rabi frequency
$\bar\Omega_\text{max} :=\nicefrac{\Omega_\text{max}}{2\pi}~\text{MHz}^{-1}$
and speedup as a sigmoid function.
To this end, we fit our simulation data to a generalized logistic function as
\begin{equation}
\label{eq:speedup}
\frac{T_\text g(\text{ARP})}{T_\text g(\text{cTQD})} = \frac{a}{1+\operatorname{e}^{-b\bar\Omega_\text{max}+c}}+d,
\end{equation}
for constants $a$, $b$, $c$ and $d$, and characterize the fit by calculating the $R^2$ score.

\section{Results}
\label{sec:results}
In this section, we present the results of our numerical analyses.
In S\ref{subsec:dipole_results} (\S\ref{subsec:quadrupole_results}), we begin by deriving generic expressions for the time-dependent coefficients of the cTQD Hamiltonian representing the dipole (quadrupole)-excitation model.
Using explicit analytical functions for these coefficients, we then simulate the operation of a cTQD CZ$_\phi$ gate and evaluate its fidelity.
Moreover, we compare these gates against CZ$_\phi$ gates realized by employing typical pulse functions, i.e., LCG and ZCHG for dipole and quadrupole excitations, respectively.
Finally in \S\ref{subsec:resource}, we compare resources required by a cTQD CZ$_\phi$ gate as compared to an adiabatic CZ$_\pi$ gate to achieve a target fidelity.

\subsection{cTQD CZ$_\phi$ gate with dipole excitation}
\label{subsec:dipole_results}
The model for a pair of interacting atoms, where each atom is dipole-excited between states $\ket1$ and $\ket{\text r}$,
is mathematically described by the cTQD Hamiltonian~$\tilde{H}^\text{d}_\text B(t)$~\eqref{eq:Hamiltonian_diabatic}.
We derive generic expressions for the time-dependent coefficients of this Hamiltonian from the adiabatic Hamiltonian~$H_\text{B}^\text{d}(t)$~\eqref{eq:Hamiltonian_adiabatic}.

We first derive the TQD Hamiltonian~$\check{H}_\text{B,eff}^\text{d}(t)$ acting on the two-dimensional subspace~$\mathscr{H}_\text{rr}$~\eqref{eq:qubitspan}.
For an effective Rabi frequency $\Omega^\text{d}_\text{eff}:=\sqrt{2}\Omega$ and an effective detuning $\Delta^\text{d}_\text{eff}:=\Delta$, the adiabatic Hamiltonian acting on $\mathscr{H}_\text{rr}$ is
\begin{align}
\label{eq:hamil_adia_eff_res}
H_\text{B,eff}^\text{d}(t)=&\frac{\Omega^\text{d}_\text{eff}}{2}\left[\ketbra{11}{+}+\ketbra{+}{11}\right]\nonumber\\
&+\frac{\Delta^\text{d}_\text{eff}}{2}[-\ketbra{11}{11}+\ketbra{+}{+}].
\end{align}
Using the TQD technique~(\S\ref{subsec:TQD}), we calculate the coefficients for the two-level control Hamiltonian~\eqref{eq:Hc_ARP} as
\begin{equation}
\label{eq:control_terms_dipole}
\Omega^\text{d}_\text{c}:=\frac{\Omega^\text{d}_\text{eff}\dot{\Delta}^\text{d}_\text{eff}-\Delta^\text{d}_\text{eff}\dot{\Omega}^\text{d}_\text{eff}}{(\Delta^\text{d}_\text{eff})^2+(\Omega^\text{d}_\text{eff})^2},\;
\theta^\text{d}:=\text{tan}^{-1}\frac{\Omega^\text{d}_\text{c}}{\Omega^\text{d}_\text{eff}}.
\end{equation}
The TQD Hamiltonian for $\mathscr{H}_\text{rr}$ is then calculated as
\begin{align}
\label{eq:HTQD_eff_dip}
\check{H}_\text{B,eff}^\text{d}(t) =&\frac{\sqrt{\left(\Omega_\text{eff}^{\text{d}}\right)^2+\left(\Omega_\text{c}^{\text{d}}\right)^2}}{2}\left[\ketbra{11}{+}+\ketbra{+}{11}\right]\nonumber\\
&+\frac{\Delta^\text{d}_\text{eff}+\dot{\theta}^\text{d}}{2}[-\ketbra{11}{11}+\ketbra{+}{+}],
\end{align}
whose dynamics effects an efficient population transfer between $\ket{11}$ and $\ket{+}$ over any integration time.

From~$\check{H}_\text{B,eff}^\text{d}(t)$, we now derive generic expressions for the time-dependent coefficients of~$\tilde{H}^\text{d}_\text B(t)$.
This is done by first obtaining~$\tilde{H}_\text{B,eff}^\text{d}(t)$ from $\check{H}_\text{B,eff}^\text{d}(t)$~\eqref{eq:checktotilde} and then performing an inversion operation on $\tilde{H}_\text{B,eff}^\text{d}(t)$~(\ref{eq:consTQDH}).
These operations yield the time-dependent Rabi frequency
\begin{align}
\label{eq:rabi_TQD_dip}
\tilde{\Omega}(t)
=&\sqrt{\frac{\left(\Omega_\text{eff}^{\text{d}}\right)^2
+\left(\Omega_\text{c}^{\text{d}}\right)^2}{2}}\nonumber\\
=&\sqrt{\Omega^2(t)+\left(\frac{\Omega(t)\dot{\Delta}(t)-\Delta(t)\dot{\Omega}(t)}{\Delta^2(t)+2\Omega^2(t)}\right)^2},
\end{align}
and detuning
\begin{align}
\label{eq:detune_TQD_dip}
\tilde{\Delta}(t)=\Delta^\text{d}_\text{eff}+\dot{\theta}^\text{d}=\Delta+\dot{\theta}^\text{d}
\end{align}
for the laser pulse in~\cref{fig:model}(a).
Using the above expressions for Rabi frequency and detuning in Eq.~(\ref{eq:Hamiltonian_dipole}), we finally derive~$\tilde{H}^\text{d}_\text B(t)$ for the two-atom system.

Having derived generic expressions for Hamiltonian coefficients, we now present numerical results on the Hamiltonian dynamics and on our fast, high-fidelity cTQD CZ$_\phi$ gate.
In the following, we state the feasible parameters for LCG pulses, compare population dynamics achieved by these pulses and their corresponding TQD pulses, and compute the pulse sequence for yielding a high-fidelity gate.

We calculate explicit expressions for the TQD pulse functions $\tilde{\Omega}^+(t)$ and $\tilde{\Delta}^+(t)$ by choosing $\Omega(t)$ and $\Delta(t)$ of the LCG form~\eqref{eq:pulses_LCG} with feasible parameters.
Using DE, we estimate feasible parameters for an adiabatic LCG pulse, of duration~$T=0.24~\upmu$s, as
\begin{equation}
\label{eq:LCGparams}
\nicefrac{\Omega_0}{2\pi}=24.92\text{ MHz},\,
\tau=0.266T,\,
\nicefrac{\Delta_0}{2\pi}=49.55\text{ MHz}.
\end{equation}
Numerical simulation of the CZ$_\pi$ gate with a double-adiabatic sequence of LCG pulses yields~$F^0_\text g=0.999$ and $T_\text g=0.48~\upmu$s.
We empirically establish that $T_\text g=0.48~\upmu$s is the minimum time for which the pulse sequence is adiabatic for the above choice of parameters and also delivers a feasible fidelity~\eqref{eq:feasibility_condition}.

\begin{figure}
    \includegraphics[width=.9\linewidth]{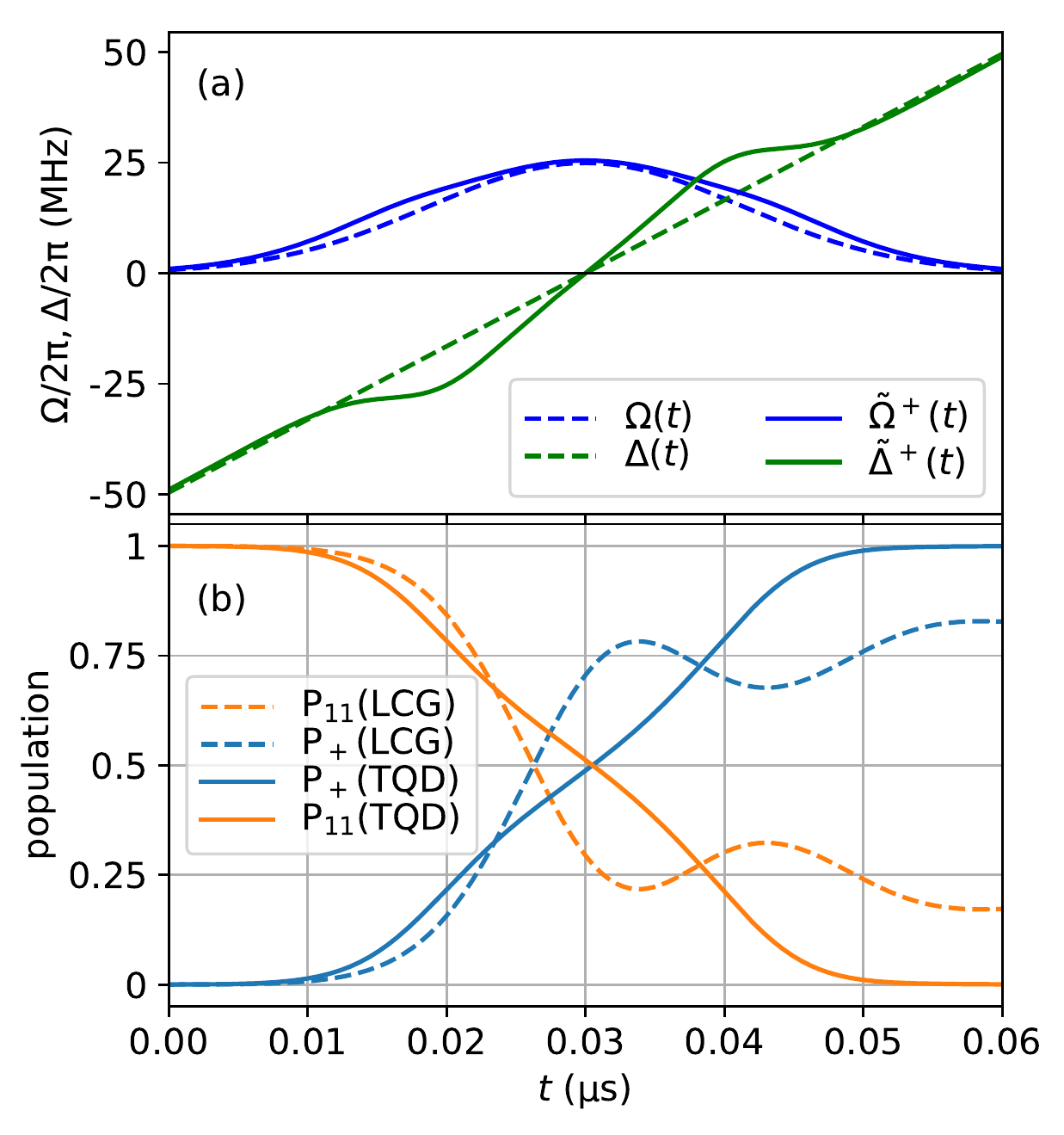}
       \caption{Pulse functions and population dynamics over $T=0.06~\upmu$s,  
      (a) Solid lines for TQD pulse functions~$\tilde\Omega^+(t)$ and $\tilde\Delta^+(t)$, dashed lines for LCG pulse functions~$\Omega(t)$ and $\Delta(t)$, with $\nicefrac{\Omega_0}{2\pi}=24.92$~MHz, $\tau=0.266T$ and $\nicefrac{\Delta_0}{2\pi}=49.55$~MHz;
    (b) Solid and dashed lines for P$_+$~(P$_{11}$) denotes the time-dependent population of the state~$\ket +$~($\ket{11}$) corresponding to the TQD and LCG pulses, respectively.}
     \label{fig:TQD}
\end{figure}

We now compare a LCG pulse against the corresponding TQD pulse, when both of these pulses are applied for one-quarter of the adiabatic LCG pulse duration, i.e.~$0.24~\upmu$s.
In~\cref{fig:TQD}(a), we plot $\Omega(t)$ and~$\Delta(t)$ corresponding to the LCG pulse of duration $T=0.06~\upmu$s and pre-determined parameters~\eqref{eq:LCGparams}.
Whereas for the TQD pulse, $\tilde\Omega^+(t)$ has a fatter tail as compared to a Gaussian function and $\tilde\Delta^+(t)$ is non-linear in time.
Furthermore, we numerically evaluate the ratio of the two pulse energies to be near unity.
In~\cref{fig:TQD}(b), we investigate the unitary dynamics of $H_\text{B}^\text{d}(t)$ ($\tilde{H}_\text{B}^\text{d}(t)$), without any parameter fluctuations, for a LCG (TQD) pulse and an initial atomic state~$\ket{11}$.
With the LCG pulse, the population~$\text{P}_+$ of the entangled state reaches 0.75 in time~$\nicefrac{T}{2}$ and changes very little, up to small oscillations, until the end of the pulse duration.
On the other hand, using the TQD pulse, $\text{P}_+$ increasing smoothly from 0 to 1 over time~$T$ and reaches $\approx1$ at $t\approx.05~\upmu$s.

\begin{figure}
    \includegraphics[width=.9\linewidth]{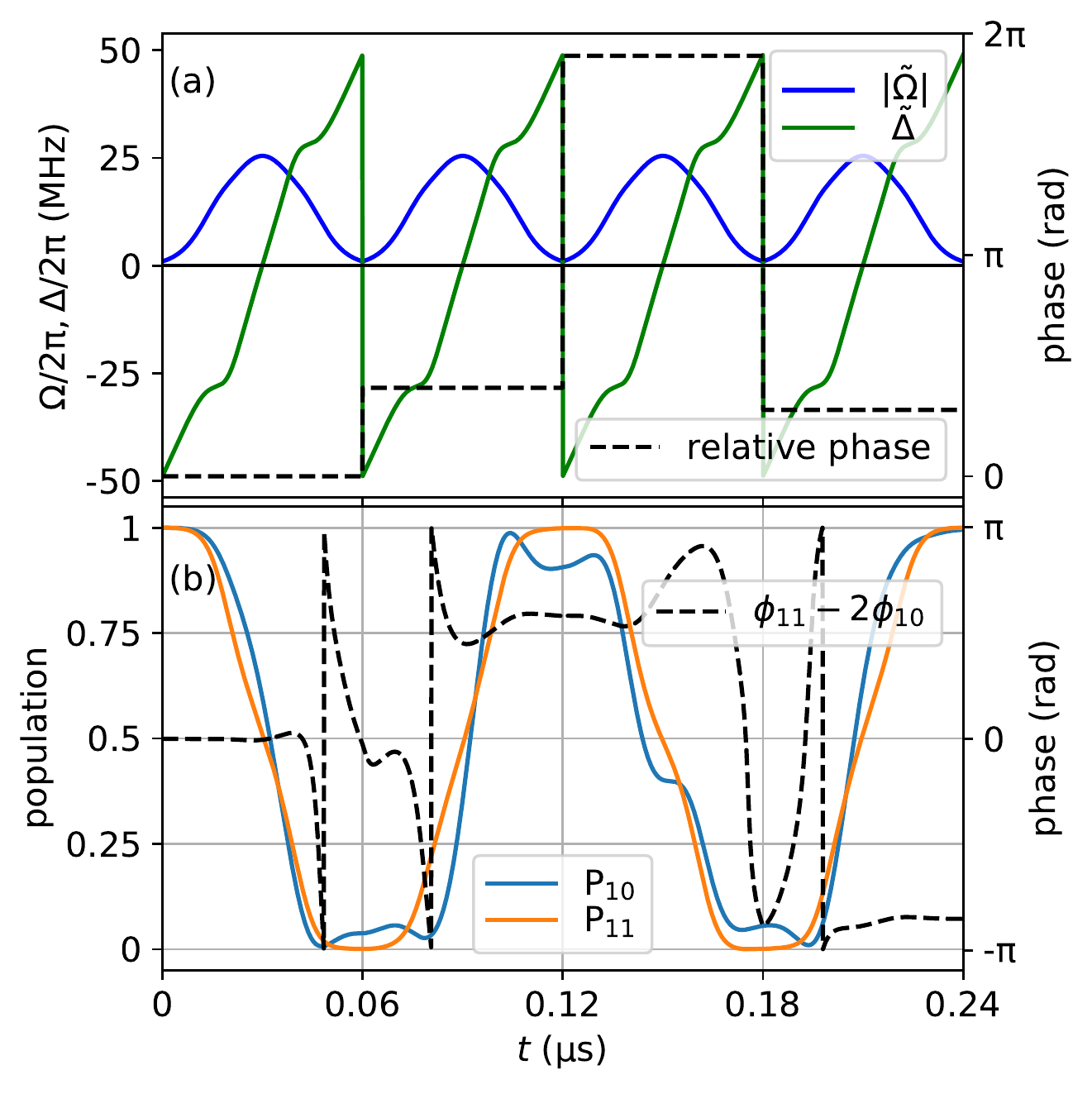}
    \caption{Pulse functions, population dynamics and phase-difference dynamics over $T_\text g=0.24~\upmu$s:
    (a) $\tilde\Omega(t)$ is a sequence of four (two pairs) time-translated $\tilde\Omega^+(t)$, with a relative phase shift of $\nicefrac{\phi_\text r}{\pi}=0.4$ between the pulses in each pair and a phase shift of $\nicefrac{\phi_\text R}{\pi}=1 .9$ between the two pairs, and~$\tilde\Delta(t)$ is a sequence of four time-translated $\tilde\Delta^+(t)$.
    (b) P$_{10}$~(P$_{11}$) is the time-dependent population of the state~$\ket{10}$~($\ket{11}$) as shown by solid lines and the dashed line shows the time-dependent phase difference~$\phi_{11}-2\phi_{10}$.
    }
     \label{fig:constrainedTQD}
\end{figure}

To implement a high fidelity cTQD CZ$_\phi$ gate in half the gate time as compared to the adiabatic CZ$_\pi$ gate, we design $\tilde{\Omega}(t)$ as a sequence of phase-shifted $\tilde\Omega^+(t)$ and $\tilde{\Delta}(t)$ as a sequence of $\tilde\Delta^+(t)$; see~\cref{fig:constrainedTQD}(a).
Based on the unitary dynamics of $\tilde{H}^\text{d}_\text B(t)$, we observe in~\cref{fig:constrainedTQD}(b) that the populations of the initial states~$\ket{10}$ and $\ket{11}$ return efficiently back to the respective states at the end of the pulse sequence.
Moreover, the phase difference~$\phi_{11}-2\phi_{10}$, after a few oscillations between $-\pi$ and $\pi$, eventually saturates at $-\pi$. 
Thus upon executing efficient population transfers and satisfying the phase relation~\eqref{eq:CZphirelation}, our gate procedure delivers $F^0_\text g=0.999$ over $T_\text g=0.24~\upmu$s, making our cTQD CZ$_\phi$ gate twice as fast as the adiabatic CZ$_\pi$ gate.

\begin{figure}
       \includegraphics[width=.9\linewidth]{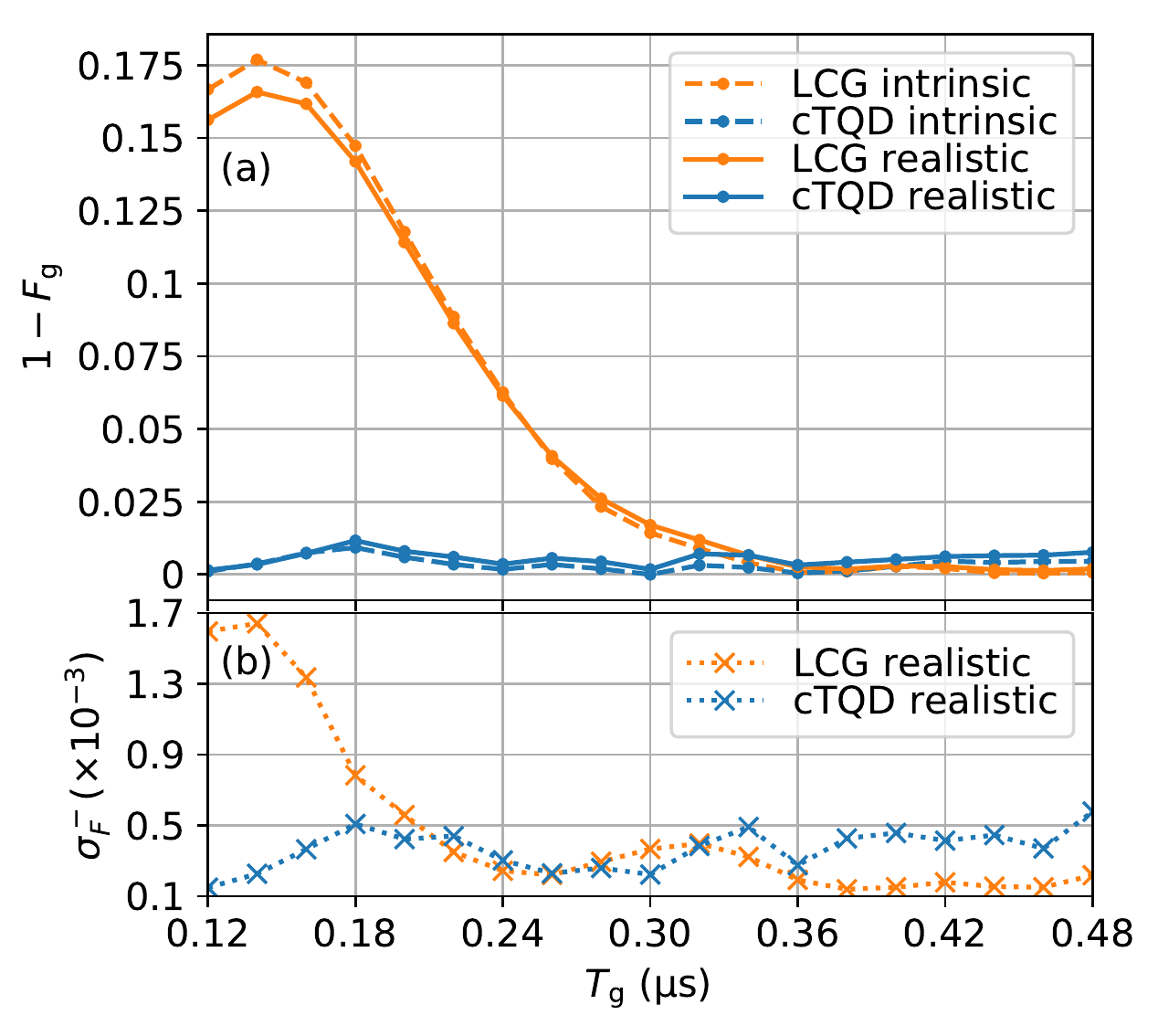}
    \caption{%
    Gate infidelity and standard error of fidelity for $T_\text g\in[0.12~\upmu\text{s}, 0.48~\upmu\text{s}]$:
    (a) Intrinsic (dashed) and realistic (solid) $1-F_\text g$ vs $T_\text g$ for LCG CZ$_\pi$ gate (orange) and cTQD CZ$_\phi$ gate (azure),
    with each point on solid curves representing the mean infidelity calculated for 100 random instances of master-equation evolution.
    (b) Each point on the dotted curves represent the standard error of fidelity calculated for these 100 fidelity estimates.
    }
     \label{fig:fidvstime}
\end{figure}

Now we investigate the variation of gate infidelity $1-F_\text g$ (both intrinsic and realistic) with changing gate time for two types of Rydberg-blockade gates, namely cTQD CZ$_\phi$ gate and LCG CZ$_\pi$ gate. 
In these simulations, we use the pre-determined values~\eqref{eq:LCGparams} for LCG parameters and consider spontaneous emission, non-zero atomic temperature, finite beam waist and fluctuations in magnetic field and laser intensity. 
The upper bound for $T_\text g$ in~\cref{fig:fidvstime}(a) is equal to the time for the adiabatic gate with LCG pulses, i.e.~$0.48~\upmu$s.
The lower bound is $4T_\text{min}$, where $T_\text{min}=0.03~\upmu$s for the TQD pulse obtained using the above LCG parameters.

Our cTQD CZ$_\phi$ gate delivers $F^0_\text g> 0.99$ and $F_\text g> 0.988$ for all $T_\text g$ between these two bounds, whereas both $F^0_\text g$ and $F_\text g$ for the LCG CZ$_\pi$ gate start to fall significantly when reducing $T_\text g$ below $0.3~\upmu$s.
We notice that the fidelity plot for cTQD CZ$_\phi$ gate have a slight oscillatory nature with respect to $T_\text g$.
Moreover, the cTQD gate fidelities have much smaller standard errors than the LCG gate fidelities for lower $T_\text g$s, whereas these errors are almost comparable for higher $T_\text g$s, as seen in~\cref{fig:fidvstime}(b).

\subsection{cTQD CZ$_\phi$ gate with quadrupole excitation}
\label{subsec:quadrupole_results}
Here we provide results on our cTQD CZ$_\phi$ gate implementation using quadrupole driving of two atoms.
First we derive expressions for the time-dependent coefficients of the cTQD Hamiltonian~$\tilde{H}^\text{q}_\text{B}(t)$~\eqref{eq:Hamiltonian_diabatic}.
Next we design a pulse sequence that yields a high-fidelity cTQD CZ$_\phi$ gate, which is twice as fast as the adiabatic CZ gate.
Finally, we show how fidelity changes with decreasing gate time in the presence and absence of imperfections and decay.

We derive an effective two-level adiabatic Hamiltonian, followed by an effective TQD Hamiltonian, with both acting on the subspace~$\mathscr H_\text{rr}$~\eqref{eq:qubitspan}.
Adiabatic elimination of state~$\ket{\text p}$ from both atoms yields a block-diagonal Hamiltonian.
After eliminating~$\ket{\text{rr}}$ according to the Rydberg-blockade condition~\eqref{eq:blockade_condition}, one of these blocks has support on only~$\mathscr H_\text{rr}$. 
This two-dimensional Hamiltonian~$H_\text{B,eff}^\text{q}(t)$ is of the same form as the effective dipole Hamiltonian~\eqref{eq:hamil_adia_eff_res}, with the corresponding time-dependent coefficients being
\begin{equation}
\label{eq:eff_pulse}
\Omega^\text{q}_\text{eff}:=-\sqrt{2}\frac{\Omega_\text{B}\Omega_\text{R}}{2\Delta_\text{B}}, \; \Delta^\text{q}_\text{eff}:=\frac{\Omega^2_\text{B}-\Omega^2_\text{R}}{4\Delta_\text{B}}. 
\end{equation}
Using the TQD technique on~$H_\text{B,eff}^\text{q}(t)$, we then construct the control terms $\Omega^\text q _\text{c}$ and $\theta^\text q$, which are similar to~Eq.~\eqref{eq:control_terms_dipole}.
Finally, we derive the effective TQD Hamiltonian
\begin{align}
\check{H}^\text{q}_\text{B,eff}(t)=&\frac{\tilde{\Omega}^\text q _\text{eff}}{2}\left[\ketbra{11}{+}+\ketbra{+}{11}\right]\nonumber\\
&+\frac{\tilde{\Delta}^\text q _\text{eff}}{2}\left[-\ketbra{11}{11}+\ketbra{+}{+}\right],
\end{align}
where the time-dependent coefficients are
\begin{equation}
\label{eq:eff_pulse_TQD}
\tilde{\Omega}^\text{q}_\text{eff}
:=\sqrt{\left(\Omega_\text{eff}^{\text{q}}\right)^2
+\left(\Omega_\text{c}^{\text{q}}\right)^2},\;
\tilde{\Delta}^\text{q}_\text{eff}:=\Delta^\text{q}_\text{eff}+\dot{\theta}^\text{q}.
\end{equation}
$H_\text{B,eff}^\text{q}(t)$ drives an efficient $\ket{11}\leftrightarrow\ket{+}$ transition in the adibatic regime, whereas $\check{H}^\text{q}_\text{B,eff}(t)$ renders this transition faster.

From~$\check{H}^\text{q}_\text{B,eff}(t)$, we now derive expressions for the time-dependent coefficients of~$\tilde{H}^\text{q}_\text{B}(t)$.
Operating under the adiabatic-elimination condition~\eqref{eq:adiabatic_elimination}, we require a constant detuning
\begin{equation}
\label{eq:detune_TQD_quad}
\tilde{\Delta}_\text B(t)\equiv\Delta_\text{B}.
\end{equation} 
Proceeding similarly to the dipole case, we first assign functions to the effective Hamiltonian~$\tilde{H}^\text{q}_\text{B,eff}(t)$ according to Eq.~\eqref{eq:checktotilde}.
Then we invert this equation,
according to Eq.~\eqref{eq:consTQDH},
to obtain time-dependent expressions for Rabi frequencies, namely, 
\begin{align}
\label{eq:rabi_TQD_quad}
\tilde{\Omega}_{\text{B}}(t)&=\sqrt{2\Delta_\text{B}\left[\sqrt{\left(\tilde{\Delta}^\text q _{\text{eff}}\right)^2+\left(\tilde{\Omega}^\text q _\text{eff}/\sqrt{2}\right)^2}+\tilde{\Delta}^\text q _{\text{eff}}\right]},\nonumber\\
\tilde{\Omega}_{\text{R}}(t)&=\sqrt{2\Delta_\text{B}\left[\sqrt{\left(\tilde{\Delta}^\text q _{\text{eff}}\right)^2+\left(\tilde{\Omega}^\text q _\text{eff}/\sqrt{2}\right)^2}-\tilde{\Delta}^\text q _{\text{eff}}\right]},
\end{align}
corresponding to the blue- and red-detuned lasers, respectively, in~\cref{fig:model}(b).
These expressions~(\ref{eq:detune_TQD_quad}) and (\ref{eq:rabi_TQD_quad}) then establish $\tilde{H}^\text{q}_\text{B}(t)$.
 
We estimate parameters for the adiabatic ZCHG pulses~\eqref{eq:pulses_ZCHG} delivering a feasible~\eqref{eq:feasibility_condition} CZ gate.
For a pair of ZCHG pulses effecting STIRAP over~$T=0.81~\upmu$s, DE yields the following feasible values for pulse parameters,
\begin{align}
\label{eq:ZCHGparams}
&\nicefrac{\Omega_\text{B0}}{2\pi}=300\text{ MHz},\,
\nicefrac{\Omega_\text{R0}}{2\pi}=300\text{ MHz}\nonumber\\
&\nicefrac{\Delta_\text{B}}{2\pi}=1762.90\text{ MHz},
\tau_\text{B}=\tau_\text{R}=0.35T.
\end{align}
Using a sequence of two STIRAP pulses, with the above parameters, we achieve a CZ gate with $F_\text g=0.996$ and $T_\text g=1.62~\upmu$s in our simulation.
We numerically demonstrate that these pulses adiabatically drive the $\ket{1}\leftrightarrow\ket{\text r}$ transition for each atom, without significantly populating the state~$\ket{\text p}$.
By eliminating $\ket{\text p}$, this transition is now an effective two-level transition. 
Furthermore, we numerically establish that $T_\text g=1.62~\upmu$s is the minimum time for which the pulses are adiabatic for the above choice of parameters.

\begin{figure}
    \includegraphics[width=.95\linewidth]{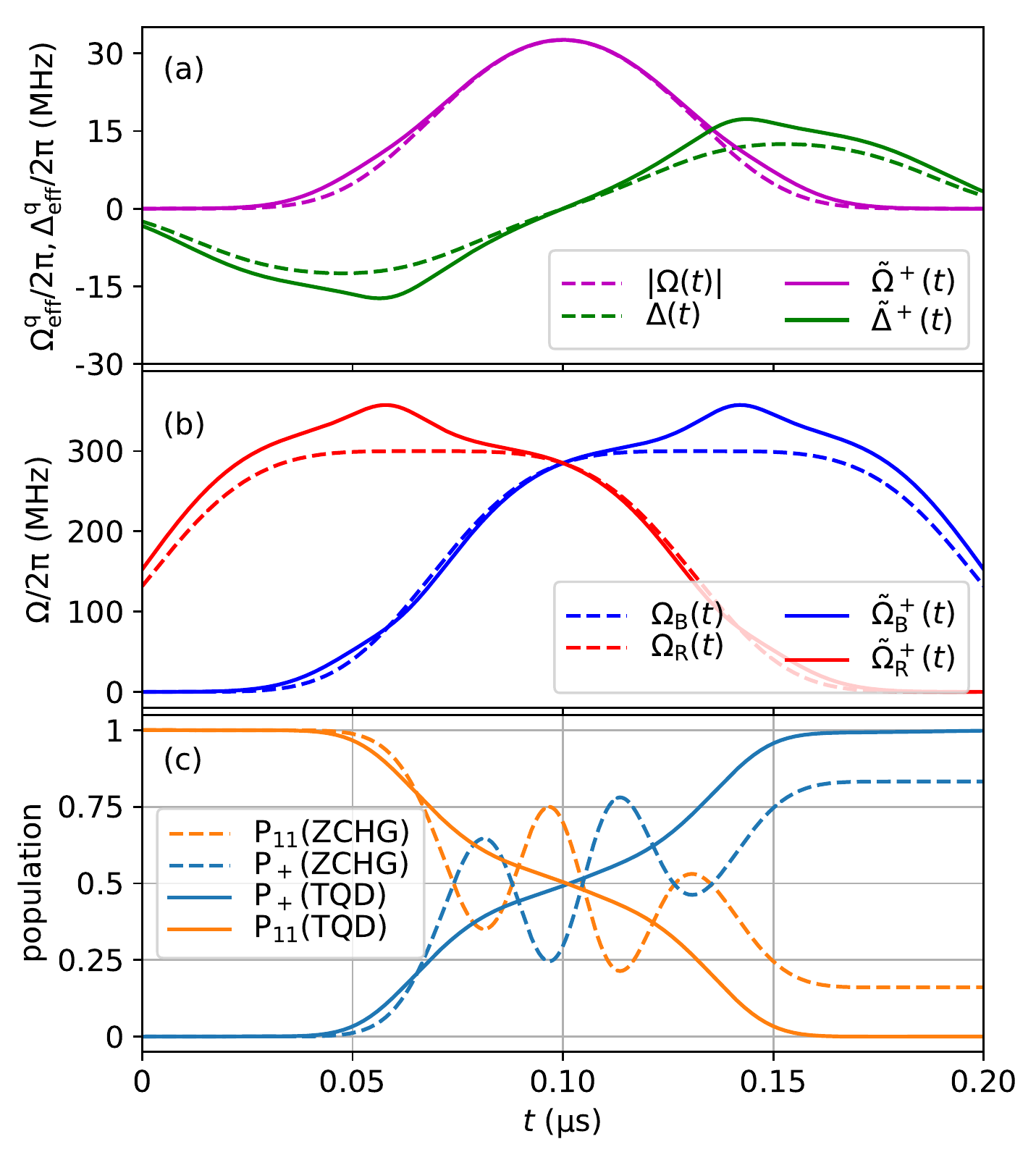}
    \caption{Laser functions and population dynamics for $T=0.2~\upmu$s and fixing other parameters as in \cref{eq:ZCHGparams}. 
    (a) Effective Rabi frequency and detuning for the pair of ZCHG pulses shown with dashed lines and for the pair of TQD pulses shown with solid lines. For simplicity, we omit subscript ``eff" and superscript ``q" in the legend.
    (b) Solid lines representing Rabi frequencies $\tilde\Omega^+_\text B(t)$ and $\tilde\Omega^+_\text R(t)$ for TQD pulses, and dashed lines representing Rabi frequencies $\Omega_\text B(t)$ and $\Omega_\text R(t)$ for ZCHG pulses.
    (c) P$_+$~(P$_{11}$) is the time-dependent population of the state~$\ket +$~($\ket{11}$) corresponding to TQD pulses (solid lines) and ZCHG pulses (dashed lines).}
     \label{fig:TQD_quad}
\end{figure}
Now we compare non-adiabatic driving by ZCHG pulses against trasitionless quantum driving of the $\ket{11}\leftrightarrow\ket +$ transition over an integration time of 0.2~$\upmu$s, which is one-quarter of the total duration for the above pair of feasible adiabatic pulses.
In \cref{fig:TQD_quad}(a), we observe that the effective Rabi frequency and detuning~(\ref{eq:eff_pulse}) for the pair of ZCHG pulses resembles Gaussian and sinusoidal functions, respectively.
The TQD technique flattens the tail of the effective Rabi frequency and add non-linear modulation to the effective detuning.
The Rabi frequencies for the TQD pulses are nearly hyper-Gaussians with a sharper and little asymmetric tip as compared to the flat top ZCHG pulses, as seen in~\cref{fig:TQD_quad}(b).
Although the peak Rabi frequencies for the two TQD pulses are significantly higher than those for their corresponding ZCHG pulses, the effective functions are almost equal.
From \cref{fig:TQD_quad}(c), we notice that the TQD pulses result in near-unity population transfer, i.e., $\text{P}_+\approx1$, at $t\approx0.16$, as compared to the 80\% saturation level using ZCHG pulses.

\begin{figure}
    \includegraphics[width=.95\linewidth]{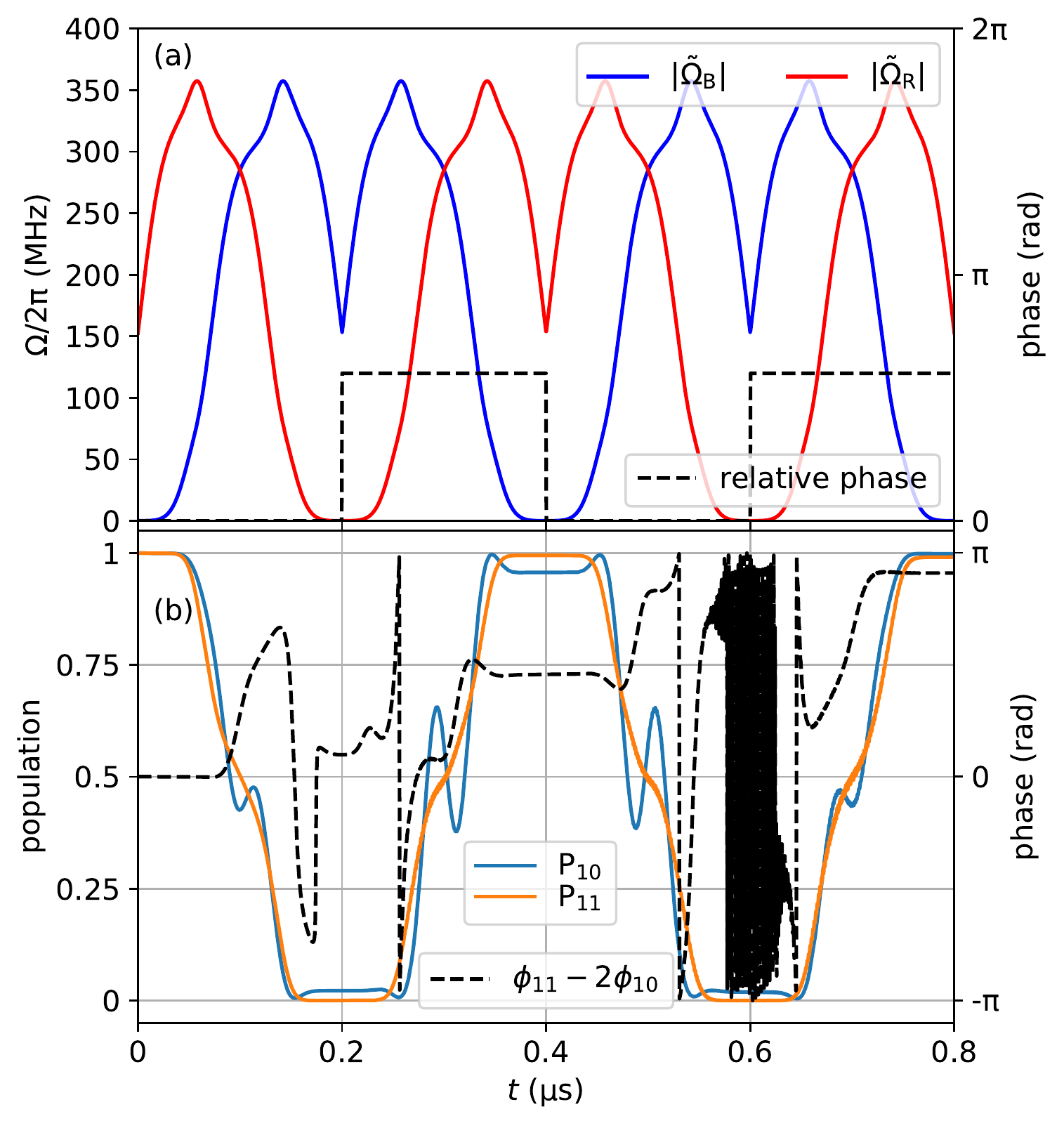}
    \caption{Pulse functions, population dynamics and phase-difference dynamics over $T_\text g=0.8~\upmu$s:
    (a) $\tilde\Omega_{\text{B,R}}$ is a sequence of four (two pairs) time-translated $\tilde\Omega^+_{\text{B,R}}(t)$, with a relative phase shift of $\nicefrac{\phi_\text r}{\pi}=0.6$ between the pulses in each pair and a phase shift of $\nicefrac{\phi_\text R}{\pi}=0$ between the two pairs, and~$\nicefrac{\tilde\Delta_\text{B}}{2\pi}=1762.90$~MHz.
    (b) P$_{10}$~(P$_{11}$) is the time-dependent population of the state~$\ket{10}$~($\ket{11}$) as shown by solid lines and the dashed line shows the time-dependent phase difference~$\phi_{11}-2\phi_{10}$.}
     \label{fig:constrainedTQD_quad}
\end{figure}
Using the pair of TQD pulses, we now construct a pulse sequence yielding a high-fidelity cTQD CZ$_\phi$ gate, which is twice as fast as the adiabatic CZ gate.
To this end, the coefficients in~$\tilde{H}^\text{q}_\text B(t)$ are designed as piecewise-smooth functions, particularly $\tilde{\Omega}_\text B(t)$ and $\tilde{\Omega}_\text R(t)$ are sequences of phase-shifted TQD pulses~(\cref{fig:constrainedTQD_quad}(a)), and $\tilde{\Delta}_\text B(t)$ is constant over the gate time.
Our estimates for relative phases ensure that the population, governed by unitary dynamics of $\tilde{H}^\text{q}_\text B(t)$, of each basis state efficiently return back to itself after time~$T_\text g=0.8~\upmu$s; see~\cref{fig:constrainedTQD_quad}(b).
Moreover, the relation~\eqref{eq:CZphirelation} between phases of basis states holds, as evident from the convergence of phase difference to 0.909$\pi$ at $T_\text g$, with high oscillations centered around $t=0.6~\upmu$s.
Consequent to the pulses satisfying the population dynmics and phase relation~(C1--C3), our predicted gate fidelity is 0.987 over a gate time of 0.8~$\upmu$s, making our gate procedure twice faster with a slight decrease in the fidelity.

We now compute fidelities for our cTQD CZ$_\phi$ gate in the presence of spontaneous emission and technical imperfections, and compare with ZCHG gate for changing gate times.
The fidelities for both intrinsic and realistic ZCHG gates oscillate rapidly with $T_\text g$, with the trough values for intrinsic fidelity gradually decreasing with increasing $T_\text g$.
In the presence of noise, fidelities drop for both higher and lower $T_\text g$s, with $\sigma_F^-$ being $10\times$ higher in higher end of $T_\text g$ as compared to the lower values.
On the other hand, cTQD gate yields $F^0_\text g>0.96$ over the whole range of $T_\text g$, with very small oscillations.
Adding noise to this gate yields lower fidelities for high $T_\text g$s, but the effect of noise is low for small $T_\text g$s.
The best estimate of fidelity is 0.975 at $T_\text g=0.24~\upmu$s.
\begin{figure}
    \centering
    \includegraphics[width=.9\linewidth]{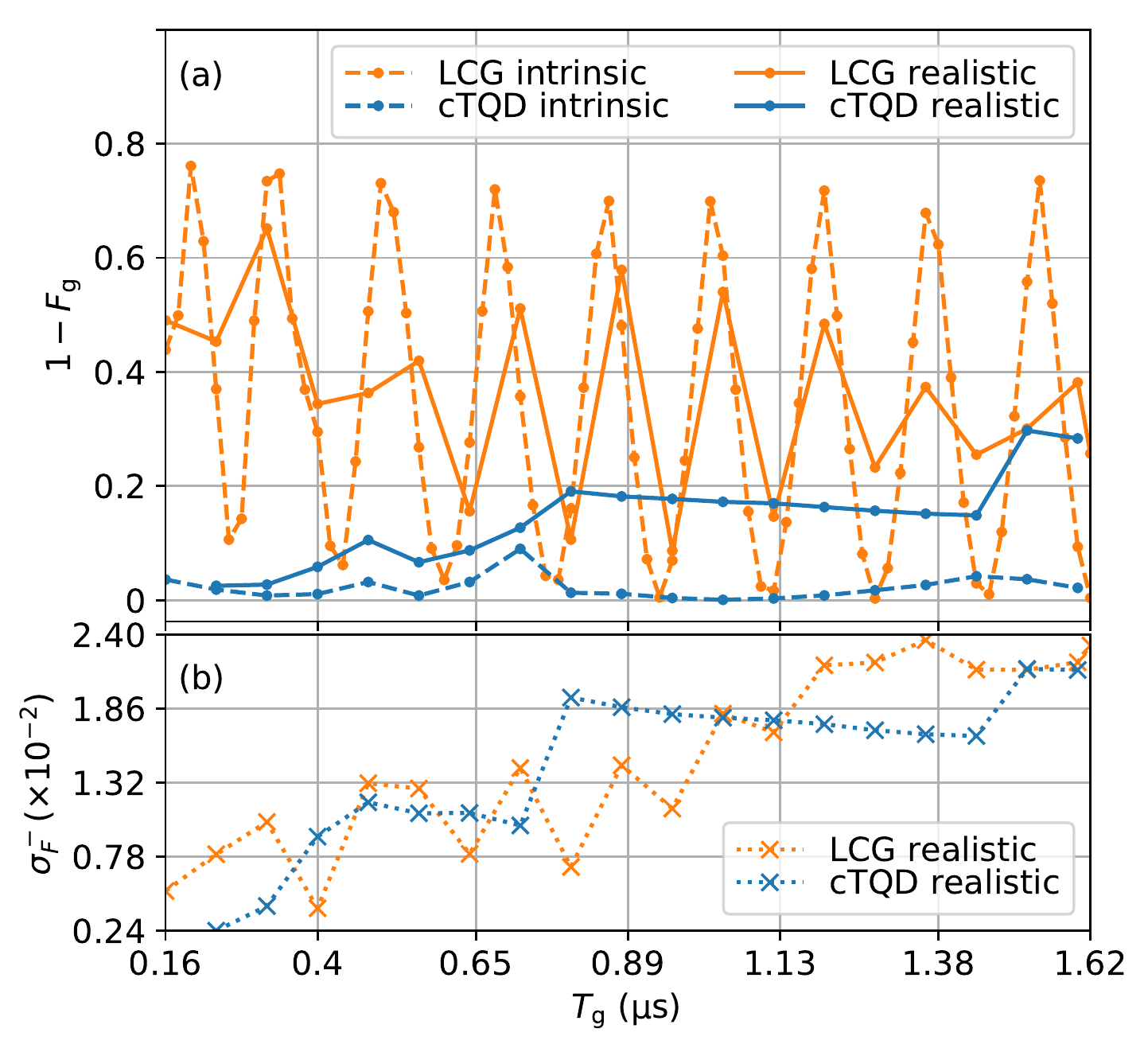}
    \caption{Gate infidelity~$1-F_\text g$ (intrinsic and realistic) vs gate time~$T_\text g$ for ZCHG pulses and cTQD pulses: intrinsic infidelity shown as dashed lines; realistic infidelity shown as solid lines with each data point and its corresponding error bar representing the sample mean and standard error of 100 MC runs.
    Chosen Parameters are $\Omega_\text{B0}/2\pi=300$~MHz, $\Omega_\text{R0}/2\pi=300$~MHz, $\Delta_\text{B}/2\pi=1762.9$~MHz.}
    \label{fig:fidvstime_quad}
\end{figure}

\subsection{Resource requirement for cTQD CZ$_\phi$ gate}
\label{subsec:resource}
We empirically calculate resources required for implementing high-fidelity CZ$_\phi$ gates by unitary evolution of adiabatic and cTQD Hamiltonians with dipole driving, 
and present these results here.
First we evaluate the scaling of $\Omega_\text{max}$ with decreasing $T_\text{g}$ for our cTQD CZ$_\phi$ gate and compare this scaling against the scaling obtained for the ARP CZ$_\pi$ gate.
Next we show how much faster the cTQD gate is over ARP gate for changing values of $\Omega_\text{max}$.
Finally, we provide a mathematical model describing the relation between speedup values and~$\Omega_\text{max}$ for our simulation data.

\begin{figure}
    \centering
    \includegraphics[width=.9\linewidth]{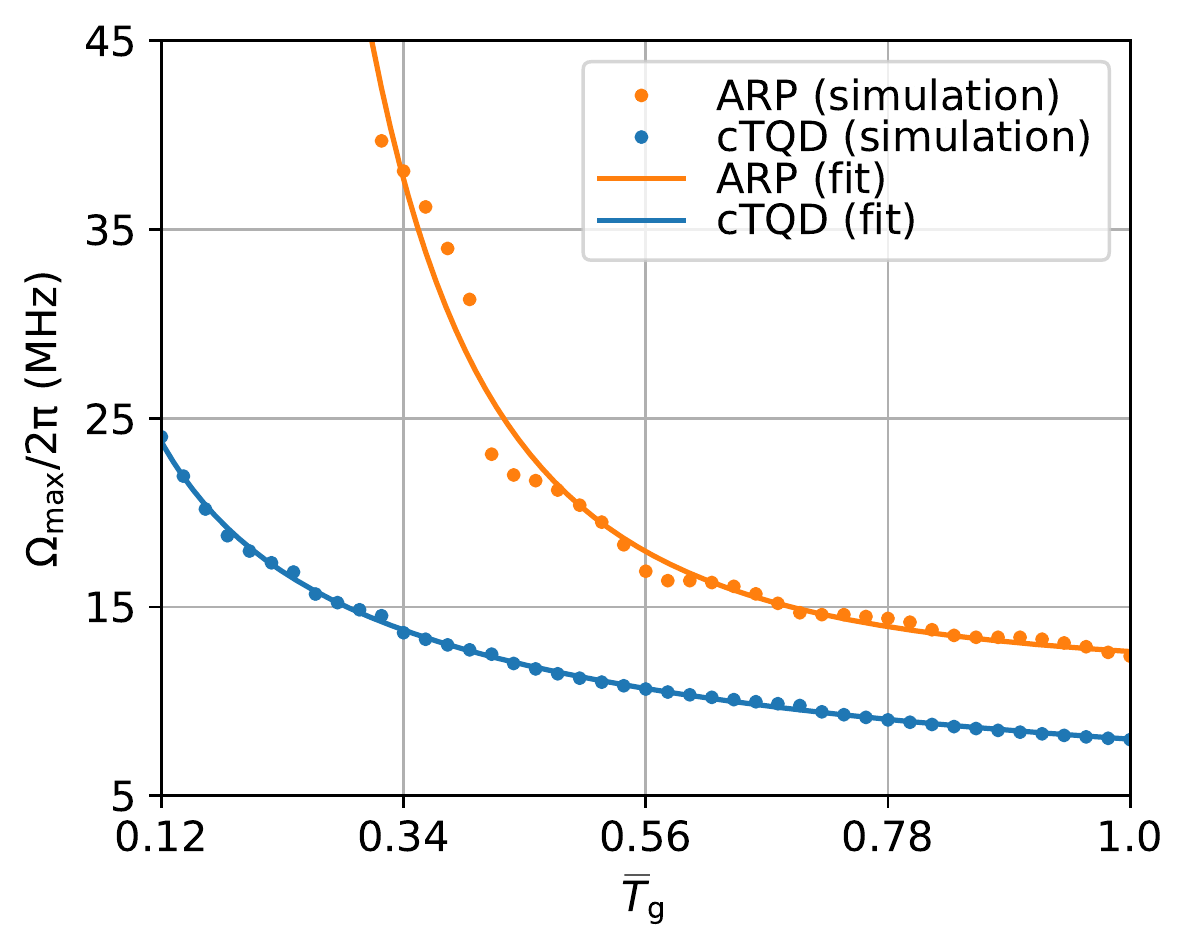}
    \caption{Scaling of maximum Rabi frequency~$\Omega_\text{max}$ with the unitless gate time~$\bar T_\text g$ for fixed $\nicefrac{\Delta_0}{2\pi}=49.55$ MHz, $\nicefrac{\tau}{T}=0.266$ and $\nicefrac{B}{2\pi}=3$ GHz: 
    Iso-fidelity plot for the ARP CZ$_\pi$ gate (orange) corresponds to $F^0_\text g>0.989$, whereas for the cTQD CZ$_\phi$ gate (azure) $F^0_\text g>0.9989$.
    Simulation data (dots) is fitted to the power function~\eqref{eq:scaling} (solid line),
    with $\nicefrac{\nu_1}{2\pi}=1.34\pm0.33~(7.18\pm0.38)$, $\nicefrac{\nu_2}{2\pi}=11.31\pm0.66~(0.82\pm0.41)$ and $p=2.76\pm0.22~(0.55\pm0.02)$ for the ARP CZ$_\pi$ (cTQD CZ$_\phi$) gate.}
    \label{fig:resources_OmVT}
\end{figure}

In~\cref{fig:resources_OmVT}, we show iso-fidelity plots for determining the scaling of $\Omega_\text{max}$ with decreasing $T_\text g$.
We vary $T_\text g$ between $0.12~\upmu$s and $1.0~\upmu$s, i.e., $\bar T_\text g\in[0.12,1.0]$, to allow for an order-of-magnitude change in gate time.
For $T_\text g<0.32~\upmu$s, ARP fails to deliver a CZ$_\pi$ gate with the target fidelity of 0.989.
The adiabatic-gate procedure exhibits a sharp fall in the required~$\Omega_\text{max}$ between gate times of $0.40~\upmu$s and $0.42~\upmu$s, whereas $\Omega_\text{max}$ for our cTQD CZ$_\phi$ gate smoothly reduces over the whole range of~$T_\text g$.
By fitting each of these two iso-fidelity plots to our assumed power function~\eqref{eq:scaling}, we evaluate $R^2$ scores of 0.97 and 0.99 corresponding to the ARP and cTQD gates, respectively.

\begin{figure}
    \centering
    \includegraphics[width=.9\linewidth]{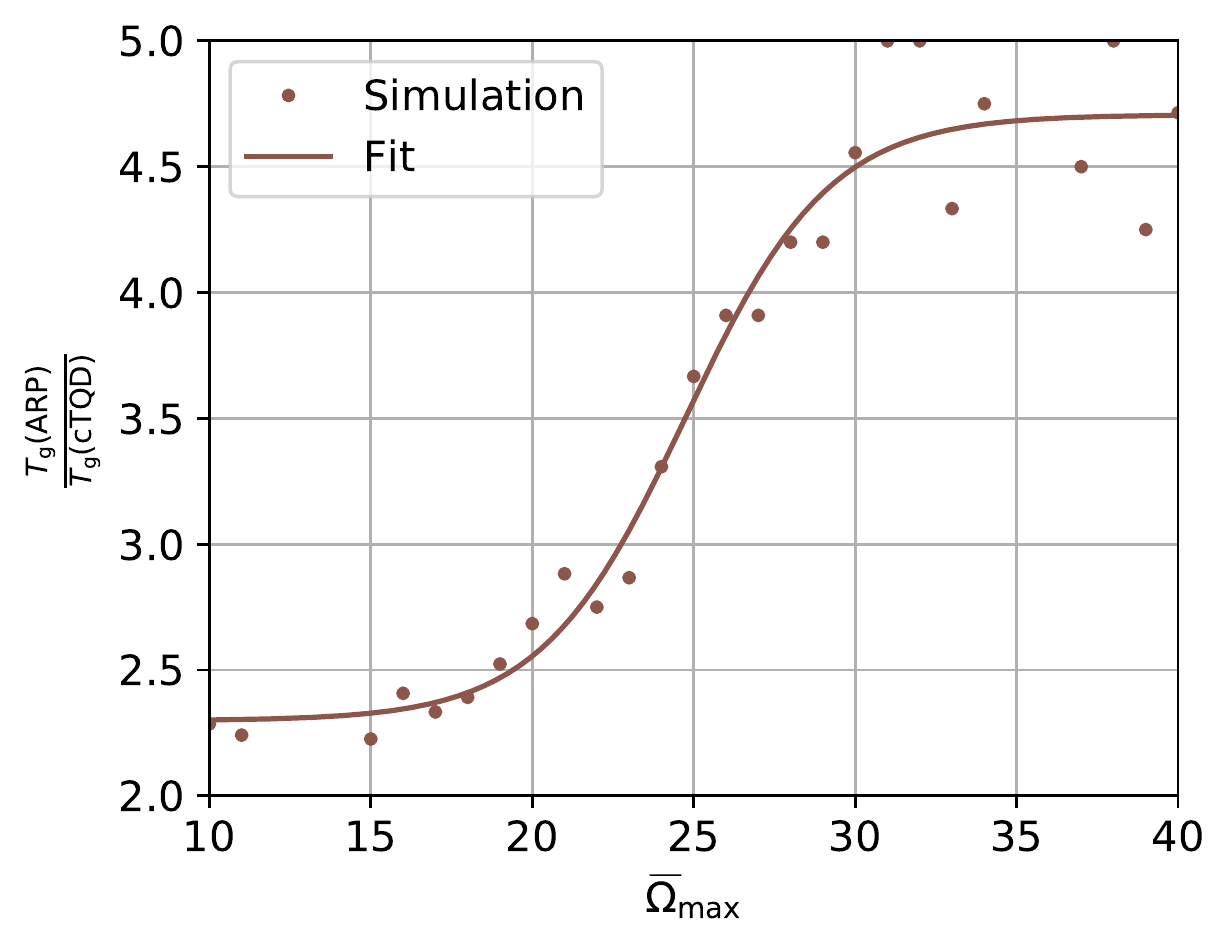}
    \caption{Trend in speedup~$\frac{T_\text g(\text{ARP})}{T_\text g(\text{cTQD})}$ with increasing values of the unitless maximum Rabi frequency~$\bar\Omega_\text{max}$ for fixed $\nicefrac{\Delta_0}{2\pi}=49.55$ MHz, $\nicefrac{\tau}{T}=0.266$ and $\nicefrac{B}{2\pi}=3$ GHz:
    Iso-fidelity plot corresponds to $F^0_\text g>0.99$ for both ARP CZ$_\pi$ and cTQD CZ$_\phi$ gates.
    Simulation data (dots) are fitted to the logistic function~\eqref{eq:speedup} (solid line) with parameters $a=2.41\pm0.15$, $b=0.45\pm0.08$, $c=11.09\pm2.05$ and $d=2.29\pm0.1$.}
    \label{fig:resources_TrVOm}
\end{figure}

We now plot the variation of speedup with $\bar \Omega_\text{max}$ in~\cref{fig:resources_TrVOm}.
Based on our results in \cref{fig:resources_OmVT}, we vary $\nicefrac{\Omega_\text{max}}{2\pi}$ between 10 MHz and 40 MHz, which translates to $\bar \Omega_\text{max}\in[10,40]$. 
We observe that the magnitudes for speedup saturate around 2.3 and 4.7 for lower and higher $\Omega_\text{max}$s, respectively.
These values match the parameters of the fitted logistic model~\eqref{eq:speedup}.
Between $\bar \Omega_\text{max}$ values of 20 and 30, we notice a rapid increase in speedup.
Additionally, fitting out simulation data to the logistic model yields a $R^2$ score of 0.96.
 
\section{Discussion}
\label{sec:discussion}
We aim to design laser-pulse functions that deliver high-fidelity, fast CZ gates for neutral atom quantum computing.
Our model includes major technical imperfections for a system of two Cs atoms simultaneously driven, either by a dipole (one-photon) process or a quadrupole (two-photon) process, between their ground and highly-excited Rydberg levels using off-resonant and broadband lasers.
The pair of Rydberg atoms experiences a strong dipole-dipole interaction, which results in a controlled-phase gate on the atomic qubits over a time~$T_\text g$.
We design these Rydberg-excitation pulses as sequences of time-dependent functions, whose expressions are derived by applying our modified TQD technique. 
Using our pulse sequences and including spontaneous emission and technical imperfections, we show that numerical integration of the master equation predicts high-fidelity CZ gates for Cs atoms.

We adapt the standard TQD technique for the interacting two-atom system, which is mathematically represented by a Hamiltonian acting on a 25-dimensional Hilbert space.
As this high-dimensional space can be decomposed into a direct sum of one- and two-dimensional subspaces under suitable approximations, we can apply the standard TQD technique independently to the Hamiltonians acting on these subspaces.
This direct application of the TQD technique results in an infeasible Hamiltonian.
Our modified TQD technique involves projecting the high-dimensional adiabatic Hamiltonian to one relevant two-dimensional subspace, applying TQD to the effective Hamiltonian of this subspace and finally lifting this TQD Hamiltonian back to a $25\times25$ non-adiabatic Hamiltonian called the cTQD Hamiltonian, which is a valid representation of our model.
Using this technique, we derive amplitude~(\ref{eq:rabi_TQD_dip},\ref{eq:rabi_TQD_quad}) and frequency~(\ref{eq:detune_TQD_dip},\ref{eq:detune_TQD_quad}) modulations for the adiabatic pulses that result in high-fidelity gates over shorter gate times.

We revisit the double-pulse adiabatic-gate procedure~\cite{SBD+20} as we need an adiabatic Hamiltonian as a starting point for constructing our cTQD Hamiltonian.
To obtain expression for this initial Hamiltonian, we describe the adiabatic transitions between ground and Rydberg levels using a LCG pulse~\eqref{eq:pulses_LCG} for the dipole model and a pair of ZCHG pulses~\eqref{eq:pulses_ZCHG} for the quadrupole model.
A global search over LCG (ZCHG) pulse parameters yields a CZ$_\pi$ (CZ) gate with intrinsic fidelity $F^0_\text g=0.999~(0.996)$.
Although the individual Rabi frequencies are high for the quadrupole transition, the effective Rabi frequency is similar to the dipole case, as evident from~\cref{fig:TQD_quad}(a).
Moreover, the gate implemented by the single-photon transition is three times as fast as the one implemented by the two-photon transition because STIRAP is inherently slower than ARP.

Using the above LCG and ZCHG pulses, we derive time-dependent TQD pulses for dipole and quadrupole driving, respectively.
In~\cref{fig:TQD}, we observe that re-distributing energy over pulse duration enables the system to reach the final state more rapidly than the linear sweep.
Intuitively, the modulation in detuning effectively cancels outs some frequency components, making the pulse narrowband and the transition more efficient. 
This analysis agree with that of Malossi et al.~\cite{MBV+13}, where they investigate population-transfer efficiencies for various pulse shapes, as opposed to reverse-engineering pulses using TQD.
For the quadrupole transition, pulse modulation by the TQD technique has the same effect on the population dynamics~(\cref{fig:TQD_quad}).
On the other hand, the LCG (ZCHG) pulse is non-adiabatic over 0.06~$\upmu$s (0.2~$\upmu$s), thus yielding an inefficient population transfer to the final states.
The oscillations in these time-dependent populations are due to transitions between the instantaneous eigenstates.

By concatenating phase-shifted, time-translated TQD pulses, we design piecewise-smooth pulse functions that  result in high-fidelity CZ$_\phi$ gates at non-adiabatic $T_\text g$s.
In Figs.~\ref{fig:constrainedTQD}(a) and \ref{fig:constrainedTQD_quad}(a),  discontinuities occuring at multiples of $\nicefrac{T_\text g}{4}$ are artifacts of this pulse-shaping method and can be smoothed out by using analytical functions~\cite{SBD+20}.
Unitary evolution generated by our cTQD Hamiltonian results in the desired population and phase dynamics, as evident from Figs.~\ref{fig:constrainedTQD}(b) and \ref{fig:constrainedTQD_quad}(b), consequently achieving high-fidelity gates.
The second pair of phase-shifted pulses aid in compensating for the loss in population of $\ket{10}$ over $\nicefrac{T_\text g}{2}$, thus further validating our pulse design procedure.
As the phase difference is approximately $\nicefrac{\pi}{2}$ at $\nicefrac{T_\text g}{2}$, a gate equivalent to $\sqrt{\text{CZ}}$ is created.
The sudden jumps in the plot for phase difference are artifacts of numerical instabilities, which can be removed by plotting with a larger time step.

We numerically investigate how $F^0_\text g$ changes with changing $T_\text g$, but keeping the maximum Rabi frequency almost constant (within 10\%).
Intuitively, we expect a steady increase in $F^0_\text g$ with increasing $T_\text g$, but in our simulation results, $F^0_\text g$ oscillates with $T_\text g$.
By calculating $F^0_\text g$ over $T_\text g<0.12~\upmu$s for the LCG CZ$_\pi$ gate~(\cref{fig:fidvstime}(a)), we observe a heavily-damped oscillation, which are far less significant as observed for the ZCHG CZ gate in~\cref{fig:fidvstime_quad}(a).
The origin of this oscillation in fidelity for the ZCHG CZ gate is the oscillation in $\phi_{11}$, with both having the same period of $0.16~\upmu$s.
The energy of the effective pulse~\eqref{eq:eff_pulse} contributes to the value of $\phi_{11}$; thus fixing this energy for decreasing $T_\text g$ removes the oscillation in fidelity.
Our cTQD CZ$_\phi$ gate yields relatively smaller fluctuations in $F^0_\text g$ with respect to $T_\text g$.

Similar to $F^0_\text g$, the realistic fidelity $F_\text g$ also oscillates with $T_\text g$.
Moreover, $F_\text g$ for the ZCHG gate drops for higher $T_\text g$s (\cref{fig:fidvstime_quad}(a)), as compared to $F_\text g$ for the LCG gate that keeps increasing (\cref{fig:fidvstime}(a)).
The adiabatic ZCHG gate is about four times slower than the adiabatic LCG gate, thus suffers from more noise as also evident from Figs.~\ref{fig:fidvstime}(b) and \ref{fig:fidvstime_quad}(b).
Our cTQD CZ$_\phi$ gate beats the LCG (ZCHG) gate at lower values for $T_\text g$s, yielding the best performance of $F_\text g=0.9985~(0.975)$ at $T_\text g=0.12~(0.24)~\upmu$s.
Whereas in the adiabatic limit, both gates perform equivalently in terms of fidelity and sensitivity to noise.

\begin{table*}[ht!]
\begin{center}
\begin{tabular}{|c|c|c|c|c|c|c|c|}
\hline
Procedure & Excitation & Implementation & Time ($\upmu$s) & $F^0_\text g$ & $F^\text{(s)}_\text g$ & $F_\text g$ & Area/$2\pi$\\
\hline
\hline
Standard & Quadrupole & Experiment~\cite{GKG+19} & 1.12 & --- & --- & 0.89 & ---\\ 
\hline
Simultaneous & Quadrupole & Experiment~\cite{GSS+22} & 0.8 & --- & --- & 0.955 & --- \\ 
\hline
Adiabatic & Dipole & Simulation~\cite{SBD+20} & 0.54 & --- & 0.999 & --- & ---\\ 
\hline
Adiabatic & Quadrupole & Simulation~\cite{SBD+20} & 1 & --- & 0.997 & --- & --- \\ 
\hline
\hline
cTQD & Dipole & Simulation & 0.12 & 0.9989 & 0.9988 & 0.9985 & 4.6\\ 
\hline
cTQD & Quadrupole & Simulation &  0.24 & 0.981 & 0.978 & 0.975 & 4.92\\ 
\hline
Adiabatic & Dipole & Simulation & 0.48 & 0.9993 & 0.9990 & 0.9980 & 14.86 \\ 
\hline
Adiabatic & Quadrupole & Simulation & 1.62 & 0.996 & 0.986 & 0.743 & 21.75\\ 
\hline
\end{tabular}
\end{center}
\caption{Comparing different gate procedures. $F^\text{(s)}_\text g$ is the fidelity in the presence of spontaneous emission from excited levels.}
\label{table:gate}
\end{table*}

We summarize performances of our cTQD CZ$_\phi$ gates, relevant experimental gates~\cite{GKG+19,GSS+22} and simulated adiabatic gates~\cite{SBD+20} in~\cref{table:gate}.
For a reasonable comparison, we only discuss experiments that use Cs atomic qubits.
Gates based on dipole excitation of Rydberg atoms, although experimentally challenging, yield highest fidelities over shortest times.
Spontaneous emission from excited levels reduce fidelity significantly for gates implemented based on quadrupole transition.
Our procedure, with the quadrupole-driving model, generates a Bell state with 50\% less infidelity than that for the state-of-the-art implementation over one-quarter gate time. 
For a detailed comparison with the adiabatic-gate procedure~\cite{SBD+20}, we report results from our simulations of adiabatic gates.
Comparing these values against cTQD gate fidelities, we infer that our procedure succeeds in making the adiabatic gates faster, but keeping $F^0_\text g$ same.

We compare the adiabatic and cTQD CZ$_\phi$ gates based on their resource requirements for preparing high-fidelity Bell states.
For simplicity, and making the reasonable approximation of a high-detuned quadrupole transition, we restrict our attention to the dipole-transition model for Rydberg excitation.
Our simulation data closely align our assumed functions relating $\Omega_\text{max}$ and $T_\text g$, as evident from their high ($>0.95$) $R^2$ values; see Figs.~\ref{fig:resources_OmVT} and \ref{fig:resources_TrVOm}.
Fixing a target fidelity, both adiabatic and cTQD gates can be sped up by increasing $\Omega_\text{max}$, but the rate of increase is different.
From~\cref{fig:resources_OmVT}, we infer that the required $\Omega_\text{max}$ for the ARP gate increases five times faster than that for the cTQD gate, when we decrease $T_\text g$ from $1~\upmu$s to $0.12~\upmu$s.
Moreover, the power factor~$p$~\eqref{eq:scaling} satisfies the inequality~\eqref{eq:ARP_condition} for the adiabatic passage, whereas the cTQD gate saturates the lower bound for this inequality by employing TQD.
In~\cref{fig:resources_TrVOm}, we notice that for a fixed value of $\Omega_\text{max}$, the cTQD gate is at least twice as fast compared to its adiabatic counterpart, but this speedup rapidly increases and eventually saturates at higher values of $\Omega_\text{max}$.
Moreover, these two plots are interrelated, i.e., one can be explained from the other.

\section{Conclusion}
\label{sec:conclusion}
Quantum computing with neutral atoms is a promising direction towards quantum advantage, with recent demonstrations of quantum algorithms~\cite{GSS+22,EKC+22}, error-correcting codes~\cite{BLS+22} and analog simulations~\cite{SLK+21,EWL+21}.
This platform bears an inherent advantage due to its unique ability to coherently control several stable qubits with the possibility of strong, long-range interactions between qubits, but state-of-the-art implementations of entangling gates yield lower fidelities than competing platforms of trapped ions and superconducting systems.
Their exists a large gap in fidelities between theoretical proposals and experimental implementations, which can be bridged by technical improvements and pulse shaping. 
In this work, we propose a new procedure for executing a two-qubit controlled phase operation that predicts high fidelity in the presence of decay from excited atomic levels and major technical imperfections.
Our results indicate that our gate procedure can pave the way for scalable quantum computing with neutral atoms.

Our procedure for constructing a Rydberg-blockade CZ gate combines our modified TQD technique with the state-of-the-art procedure for implementing CZ gates on atomic qubits~\cite{LKS+19}.
Our symmetric CZ gate is executed between two trapped atoms by simultaneously driving both atoms between their ground and Rydberg levels using focused pulses on individual atoms.
We consider both one- and two-photon processes for the ground to Rydberg excitation, which are feasible with current experimental setups~\cite{GSS+22,MJL+21}. 
Furthermore, we incorporate major sources of technical imperfections in our model, namely Doppler dephasing,
atomic-position fluctuation and laser-intensity fluctuation, and make our simulations realistic by using actual values of these imperfections~\cite{GSS+22}. 
Our procedure reduces gate errors originating from atom loss and is feasible for implementing CZ gates on large arrays of trapped atoms.

By following our gate procedure, we design time-dependent functions for the Rydberg-excitation lasers that result in high-fidelity CZ gates.
These pulse functions are derived from adiabatic pulses by using our modified TQD technique.
We analysed the impact of this technique by comparing pulse shapes and Hamiltonian dynamics.
Our results show that re-distributing energy by modifying pulse shapes according to the TQD technique speeds up quantum processes.
Additionally, we design a sequence of such modified pulses, with relative phase shifts, to achieve a fast, high-fidelity CZ gate.

We estimate CZ gate fidelities for Cs atom qubits by numerically integrating the quantum master equation. 
In the presence of spontaneous emission from excited Cs levels and major technical imperfections, our gate procedure deliver CZ gates with fidelities 0.9985 and 0.975 for the one- and two-photon excitation models, respectively.
Although one-photon excitation model leads to higher fidelity, its experimental realization is challenging and needs improvements.
As compared to adiabatic CZ gates, our gates yield $0.1\times$ the infidelity in $0.15\times$ the gate time requiring $0.3\times$ the pulse energy.
The infidelity of the state-of-the-art experimental CZ gate is $2\times$ and the gate time is $3.3\times$ than that of our simulated CZ gate with the same excitation model.
Our gates are robust against Doppler dephasing of the Rydberg level and changes in the duration of the Rydberg-excitation pulses
Moreover, our gates are less sensitive to thermal fluctuations of atoms and intensity fluctuations of pulses as compared to adiabatic gates.

Our modified TQD technique can be used for speeding up adiabatic dynamics in high-dimensional Hilbert spaces.
Although our gate procedure is tailored for Rydberg-blockade gates, our approach of combining TQD with existing gate procedures can be explored for different gates in other quantum computing platforms. 
Thus our work introduces a powerful pulse-shaping technique in the quantum control toolbox.
Similar to the standard TQD technique, our pulse functions are not optimal and their performnce efficiencies are limited by the choice of the adiabatic pulses used to derive them.
An interesting future direction is to use optimized STIRAP pulses~\cite{VKV09}.
Another possibility is to derive closed-form expressions~\cite{LKS+19} for the two phase-shift parameters in our pulse sequences and exploit the interplay between pulse parameters to improve gate performance.


\acknowledgments
In the spirit of reconciliation, we acknowledge that we live, work and play on the traditional territories of the Blackfoot Confederacy (Siksika, Kainai, Piikani), the Tsuut’ina, the \^{I}y\^{a}xe Nakoda Nations, the M\'{e}tis Nation (Region 3), and all people who make their homes in the Treaty 7 region of Southern Alberta.
This project is supported by Government of Alberta and Natural Sciences and
Engineering Research Council of Canada (NSERC). 
AD acknowledges financial support from Mitacs and computational support from Compute Canada Calcul Canada.


\bibliography{adiabatic.bib,tqd.bib}

\begin{appendix}
\section{Mathematical precepts}
\label{app:math}
Here we discuss the decomposition of the 25-dimensional Hamiltonian $\tilde{H}_\text{B}^{\text e}(t, \bm r)$~\eqref{eq:Hamiltonian_diabatic} into a direct sum of one- and two-dimensional Hamiltonians, and
the relevant matrix transformations for this decomposition.
In this regard, we define an operator that maps the 25-dimensional Hamiltonian to an effective two-dimensional Hamiltonian.
The mapping involves matrix transformations which we express as orthogonal matrices for permutation and basis transformation, and idempotent-symmetric matrix for projection.
The decomposition and mapping of the time-dependent Hamiltonians are defined at each time~$t$ during the dynamics.  

\subsection{Transformations}
We first define of the transformations and operators that we use in this section.
For a Hilbert space~$\mathscr{H}$, $\mathcal{B}(\mathscr{H})$ denotes the space of bounded linear operators acting on $\mathscr{H}$.
In our case, the transformations, which are basis change and projection, of these operators are represented as conjugating channels.
The transformations of the Hamiltonian operators are effected by operators representing the corresponding transformation.
\begin{definition}
A conjugating-channel transformation is
\begin{equation}
\label{eq:conj_chan}
  \mathcal{C}:
  \mathcal{B}(\mathscr{H})\times\mathcal{B}(\mathscr{H})
  \to\mathcal{B}(\mathscr{H}):
  \left(A,B\right)\mapsto ABA^\dagger
\end{equation}
for any $A\in\mathcal{B}(\mathscr{H})$,
and we use the notation
$\mathcal{C}_A(B)=ABA^\dagger$. 
Here, $A$ is the transformation operator and $B$ is the Hamiltonian operator.
\end{definition}

The decomposition of the Hilbert space~$\mathscr{H}$ into two subspaces~$\mathscr{H}_1$ and $\mathscr{H}_2$ is represented as a direct sum
\begin{equation}
\label{eq:decompose}
   d =  d_1 \oplus d_2,
\end{equation}
where $d,d_1$ and $d_2$ are the dimensions of $\mathscr{H}$, $\mathscr{H}_1$ and $\mathscr{H}_2$, respectively.
A projection on~$\mathscr{H}$ is a conjugating-channel transformation~\eqref{eq:conj_chan}, which is executed by a real-valued idempotent operator~$\pi$ with range $\mathscr{H}_1$ and kernel $\mathscr{H}_2$. 
For the special case of $B$ being a block-diagonal matrix with two blocks $H_1\in\mathcal{B}(\mathscr{H}_1)$ and $H_2\in\mathcal{B}(\mathscr{H}_2)$,
\begin{equation}
\label{eq:projection}
\mathcal{C}_\pi(B)=\pi B \pi = H_1 \oplus \mathbb0_{d_2}=:B'
\end{equation}
for~$\mathbb0_{d_2}$ the $d_2\times d_2$ null matrix. Thus, the operation~$\pi$ maps a $d\times d$ matrix~$B$ to a $d_1\times d_1$ matrix~$H_1$, padded with a null matrix.

The inverse of a conjugating-channel transformation satisfies
\begin{equation}
\label{eq:inversetrans}
  \mathcal{C}^{-1}_A(B) = A^{-1}BA^{\dagger^{-1}}
  =A^{-1}BA^{-1^{\dagger}}
\end{equation}
for any invertible~$A$.
Particularly, for orthogonal and real-valued~$A$, representing a basis-change operator,
\begin{equation}
\label{eq:orthoreal}
\mathcal{C}^{-1}_A(B)
=A^\top BA.
\end{equation}
For a projection operator~$\pi$ satisfying~\eqref{eq:projection}, we define a unique lift operator~$\rotpi$ that executes an inverse projection transformation such that
\begin{equation}
\label{eq:idemreal}
\mathcal{C}^{-1}_{\pi} (B') =\rotpi B' \rotpi \equiv B,
\end{equation}
where $B$ and $B'$ are block-diagonal matrices.

\subsection{Decomposition}
For our purpose, we first permute the matrix representation of~$\tilde{H}^\text{e}_\text{B}(t)$ into a block-diagonal structure by the conjugating-channel transformation~$\mathcal{C}_{P_1}\left(\tilde{H}^\text{e}_\text{B}(t)\right)$, for a orthogonal and real-valued $P_1$.
This transformation makes $\tilde{H}^\text{e}_\text{B}(t)$ a direct sum of $9\times9$ and $16\times16$ Hamiltonian matrices, and thus decomposing $\mathscr{H}$ as
\begin{equation}
\label{eq:block_diagonal}
   25 =  5\otimes5 \overset{}{\approx} 9\oplus16.
\end{equation}
This block diagonalization is justified by not creating any coupling between levels~$\{\ket0,\ket1,\ket{\text r}\}$, whose tensor products with themselves span the 9-dimensional Hilbert space~$\mathscr{H}_\text{CZ}$, and the other levels~$\{ \ket{\text g},\ket{\text p}\}$.
In the quadrupole case, this decoupling is not strictly true because~$\ket{\text p}$ is employed virtually for $\ket1\leftrightarrow\ket{\text r}$ coupling, but we eliminate~$\ket{\text p}$ by adiabatic methods and thus justify~(\ref{eq:block_diagonal}) for the quadrupole case.

After a projection transformation~$\mathcal{C}_{\pi_1}$ of the block-diagonal Hamiltonian, we obtain a $9\times9$ Hamiltonian matrix~$\tilde{H}^\text{e}_\text{B,CZ}(t)$ by
\begin{equation}
\mathcal{C}_{\pi_1}\circ\mathcal{C}_{P_1}\left(\tilde{H}^\text{e}_\text{B}(t)\right) = \pi_1P_1\tilde{H}^\text{e}_\text{B}(t)P_1^\top \pi_1 = \tilde{H}^\text{e}_\text{B,CZ}(t) \oplus\mathbb0_{16}.
\end{equation}
This $9\times9$ matrix represents the unitary dynamics of~$\tilde{H}_\text{B}^\text{e}(t)$ exactly for the dipole case and approximately for the quadrupole case.

We now perform a permutation transformation~$\mathcal{C}_{P_2}$ of $\tilde{H}^\text{e}_\text{B,CZ}(t)$ to derive a block-diagonal structure according to the decomposition
\begin{equation}
\label{eq:P2}
9 = 3\otimes3 \overset{}{=} 1\oplus2\oplus2\oplus4
\end{equation}
of $\mathscr{H}_\text{CZ}$.
The singleton arises because there is no coupling between $\ket{00}$ and other levels. 
The existence of the two-dimensional sectors is justified if only one of the two atoms is in $\ket1$ and hence driven by the Rydberg excitation laser to $\ket{\text r}$. 

The four-dimensional sector, which represents the case when both atoms are driven, is spanned by a basis set
\begin{equation}
\{\ket{11}, \ket+, \ket{\text{rr}},
\ket-\},
\end{equation}
obtained from the standard basis~$\{\ket{11}, \ket{1\text r}, \ket{\text r 1}, \ket{\text{rr}}\}$ by transformation~$\mathcal{C}_Q$.
In this new basis, the block-diagonal structure~\eqref{eq:P2} further decomposes as
\begin{align}
9=&1\oplus2\oplus2\oplus4 \approx  1\oplus2\oplus2\oplus2\oplus1\oplus1 \nonumber\\
= & \underbrace{\operatorname{span}\{\ket{00}\}}_{\mathscr{H}_0} \oplus
\underbrace{\operatorname{span}\{\ket{01},\ket{0\text{r}}\}}_{\mathscr{H}_{0\text{r}}} \oplus
\underbrace{\operatorname{span}\{\ket{10},\ket{\text{r}0}\}}_{\mathscr{H}_{\text{r}0}} \nonumber \\
&\oplus\underbrace{\operatorname{span}
	\{\ket{11}, \ket{+}\}}_{\mathscr H_\text{rr}} 
\oplus \operatorname{span}\{\ket{-}\} \oplus \operatorname{span}\{\ket{\text{rr}}\}.
\end{align}
where $\ket-$ is a dark state and $\ket{\text{rr}}$ is decoupled from the rest of the states by adiabatic elimination under Rydberg-blockade condition~\eqref{eq:blockade_condition}.

Finally, we define an effective two-level Hamiltonian~$\tilde{H}_\text{B,eff}^\text{e}(t)$  as a final projection~$\mathcal{C}_{\pi_2}$ of $\tilde{H}_\text{B}^\text{e}(t)$ onto $\mathscr H_\text{rr}$.
Thus, the whole operation of mapping the $25\times25$ matrix~$\tilde{H}_\text{B}^\text{e}(t)$ to a $2\times2$ matrix~$\tilde{H}_\text{B,eff}^\text{e}(t)$ is represented by the conjugating-channel transformation
\begin{equation}
\tilde{H}_\text{B,eff}^\text{e}(t, \bm r) \oplus \mathbb0_{23}
=\mathcal{C}_R\left(\tilde{H}_\text{B}^\text{e}(t, \bm r)\right)
\equiv R\tilde{H}_\text{B}^\text{e}(t, \bm r)R^\top ,
\end{equation}
for a real-valued operator
\begin{equation}
\label{eq:Rdef}
R:=\pi_2QP_2\pi_1P_1.
\end{equation}
$\tilde{H}_\text{B,eff}^\text{e}(t)$ drives the transition between $\ket{11}$ and $\ket{+}$ and is primarily responsible for entanglement.

Although the dynamics of the two-atom system can be broken up into independent singleton and qubit evolutions, the initial state~$\ket{\psi_0}$~\eqref{eq:initial_state} has support over all four Hilbert-space sectors~$\mathscr{H}_0$, $\mathscr{H}_{0\text{r}}$, $\mathscr{H}_{\text{r}0}$ and $\mathscr{H}_{\text{r}\text{r}}$.
The~$\ket{00}$ component of the initial state,
i.e., the projection of the initial state onto~$\mathscr{H}_0$,
undergoes trivial (i.e., identity~$\mathds1$) evolution,
so the~$\ket{00}$ coefficient of the instantaneous state has a constant magnitude during the evolution.
As for the other three components of the initial state, their coefficients in the instantaneous state evolve according to the two-level Hamiltonians obtained by projecting~$\tilde{H}_\text{B}^\text{e}(t)$ onto~$\mathscr{H}_{0\text{r}}$, $\mathscr{H}_{\text{r}0}$ and $\mathscr{H}_{\text{r}\text{r}}$, respectively.
\end{appendix}

\end{document}